\documentclass[twocolumn,showpacs,aps,prd]{revtex4}

\usepackage{dcolumn}
\usepackage{amsmath}
\usepackage{graphicx}
\usepackage{epsfig}

\usepackage{color}
\usepackage{graphics, graphicx}
\DeclareGraphicsExtensions{.eps}
\usepackage{epsfig}	

\definecolor{purple}{rgb}{0.5,0,0.5}
\definecolor{darkgreen}{rgb}{0.0,0.7,0.2}
\definecolor{forest}{rgb}{0.1,0.3,0.2}
\definecolor{darkred}{rgb}{0.7,0.0,0.2}
\definecolor{darkcyan}{rgb}{0.,0.2,0.8}
\definecolor{orange}{rgb}{0.7,0.2,0.0}

\def\funct{function}

 \newcommand{\recow}{\mbox{${w}$}}
 \newcommand{\ctl}{\mbox{${\cos \theta_\ell}$}}
 \newcommand{\ctv}{\mbox{${\cos \theta_V}$}}
 \newcommand{\angchi}{\mbox{${\chi}$}}

 \newcommand{\ffH}{\mbox{$H$}}

\newcommand{\stl}{\mbox{${\sin \theta_\ell}$}}
\newcommand{\stv}{\mbox{${\sin \theta_V}$}}

\newcommand{\rone}{ {\mbox{$R_1$}}}
\newcommand{\rtwo}{{\mbox{$R_2$}}}

\def\rhosq{{\mbox{$\rho^2$}}}

\def\babar{BaBar}

\def\comb{combinatorial}
\def\Comb{Combinatorial}
\def\combb{combinatorial background}

\def\histo{histogram}
\def\diml{dimensional}

\def\sbr{sideband region}

\def\coef{coefficient}

\def\bkgd{background}

\def\pking{peaking}

\def\Bkgd{Background}

\def\dist{distribution}

\def\statl{statistical}

\def\syst{systematic}
\def\Syst{Systematic}
\def\systerr{systematic error}

\def\likeld{likelihood}
\def\param{parameter}
\def\paramzn{parameterization}

\def\Param{Parameter}

\def\diff{difference}

\def\recod{reconstructed}

\def\normzn{normalization}
\def\FF{form factor}
\def\ff{form factor}

\def\reco{reconstruct}

\def\Ndata{{N_{\rm data}}}

\def\bef{\begin{figure}}
\def\eef{\end{figure}}

\newcommand{\bsl}{\begin{slide}}
\newcommand{\esl}{\end{slide}}

\def\Btodstarenu{\mbox{$B^{0}\rightarrow D^{*-} e^+ \nu_e$}}
\def\Bbartodstarlnu{\mbox{${\bar B}^{0}\rightarrow D^{*+}\ell^{-} \nubar$}}

\def\Bbartodstarenu{\mbox{${\bar B}^{0}\rightarrow D^{*+}e^{-} \nuebar$}}

\def\Bbartodslnu{\mbox{${\bar B} \rightarrow D^{*}\ell \nubar$}}

\def\btodstarlnu{\mbox{${ B^0 \rightarrow D^* \ell  \nu}$ } }

\def\btodstarenu{\mbox{${ B^0 \rightarrow D^* e  \nu}$ } }

\def\h0{H_{0}(q^2)}

\def\mdstar{M_{D^{*}}}

\def\pip{\pi^{+}}

\def\cosby{\mbox{$\cos\theta_{BY}$}}

\def\m2miss{M^{2}_{\rm miss}}

\def\mb{M_B}

\def\momdstr{\mbox{$\wp_{D^{*}}$}}

\def\vdstr{v_{D^{*}}}

\def\upsfs{\mbox{$\Upsilon(4{\rm S})$}}

\def\be{\begin{equation}}
\def\ee{\end{equation}}
\def\bea{\begin{eqnarray}}
\def\eea{\end{eqnarray}}

\def\bess{\begin{equation*}}
\def\eess{\end{equation*}}

\def\pstarl{\mbox{$p^*_\ell$}}

\newcommand{\rp}{\right)}
\newcommand{\lp}{\left(}

\newcommand{\lb}{\left[}
\newcommand{\rb}{\right]}

\newcommand{\eps}{\epsilon}

\newbox\charbox
\newbox\slabox
\def\s#1{{      
        \setbox\charbox=\hbox{$#1$}
        \setbox\slabox=\hbox{$/$}
        \dimen\charbox=\ht\slabox
        \advance\dimen\charbox by -\dp\slabox
        \advance\dimen\charbox by -\ht\charbox
        \advance\dimen\charbox by \dp\charbox
        \divide\dimen\charbox by 2
        \raise-\dimen\charbox\hbox to \wd\charbox{\hss/\hss}
        \llap{$#1$}
}}

\newcommand{\as}{\mbox{$\alpha_{s}$}}
\newcommand{\asq}{\mbox{$\alpha_{s}^2\ $}}

 \newcommand{\benn}{\begin{displaymath}}
 \newcommand{\eenn}{\end{displaymath}}

 \newcommand{\bes}{\begin{displaymath}}
 \newcommand{\ees}{\end{displaymath}}

 \newcommand{\beas}{\begin{eqnarray*}}
 \newcommand{\eeas}{\end{eqnarray*}}

\newcommand{\bean}{\begin{eqnarray*}}
 \newcommand{\eean}{\end{eqnarray*}}

\newcommand{\dmu}{D_\mu}

\newcommand{\rar}{\mbox{$\rightarrow$}}

\newcommand{\piplus}{\mbox{$\pi^+$}}
\newcommand{\piminus}{\mbox{$\pi^-$}}

\newcommand{\eminus}{\mbox{$e^-$}}
\newcommand{\eplus}{\mbox{$e^+$}}

\def\epem{\eplus\eminus}

\newcommand{\nuebar}{\overline{\nu}_e}

\newcommand{\nue}{\mbox{$\nu$}}

\newcommand{\vcb}{\mbox{$V_{\rm cb}$}}

\newcommand{\ubar}{\overline{u}}
\newcommand{\dbar}{\overline{d}}

\newcommand{\nubar}{\overline{\nu}}

\newcommand{\sbar}{\overline{s}}

\newcommand{\dstar}{\mbox{$D^{*}$}}

\newcommand{\Bbar}{\overline{B}}

\newcommand{\Dbar}{\overline{D}}

\newcommand{\bi}{\begin{itemize}}
\newcommand{\ei}{\end{itemize}}

\newcommand{\Btodslnu}{\mbox{ $B \rar D^* \ell \nu$} }

\newcommand{\ifb}{\mbox{ $\rm fb^{-1}$}}

\newcommand{\Dzero}{\mbox{$D^0$}}

\newcommand{\Dstar}{\mbox{$D^*$}}
\newcommand{\Dstarplus}{\mbox{$D^{*+}$}}

\newcommand{\Dst}{\mbox{ $D^*$} }

\newcommand{\deltam}{\mbox{$\delta m$}}

\newcommand{\Deltam}{\mbox{$\Delta m$}}

\newcommand{\order}{\mbox{$\mathcal O$}}

\newcommand{\vsp}{\vspace{.5cm}}
\newcommand{\vsps}{\vspace{.5cm}}

\newcommand{\Bzero}{\mbox{$B^0$} }

\newcommand{\Bzerobar}{\mbox{$ \overline{B}^0$}}

\def\kv{kinematic variable}

\def\KV{Kinematic Variable}

\def\ndof{\mbox{$n_{\rm dof}$}}

\newcommand{\Bz}{$B^0$}

\newcommand{\spi}{\mbox{$\pi_s$}}

\def\lqcdomx{ { \lp {\Lambda_{\rm QCD} \over m_x }\rp  }}

\newcommand{\defineCat}[1]{%
\expandafter\def\csname #1\endcsname{{\ttfamily #1}\xspace}}
\defineCat{svtDch}
\defineCat{Dmode}
\defineCat{tagCat}
\defineCat{angCut}
\defineCat{leptID}
\defineCat{onOffRes}
\defineCat{peakGroup}
\defineCat{yields}

\def\dstst{\mbox{$D^{**}$}}

\def\Dstst{\mbox{$D^{**}$}}

\def\resln{resolution}

\def\thetav{\theta_V}
\def\thetal{\theta_\ell}

\def\subsec{\subsection}
\def\subsubsec{\subsubsection}

\def\ssec{\subsection}
\def\sssec{\subsubsection}

\def\aone{\mbox{$A_1$}}
\def\atwo{\mbox{$A_2$}}

\def\bcenter{\begin{center}}
\def\ecenter{\end{center}}

\def\gof{goodness-of-fit}

\def\Appx{Appendix}

\def\aone{\mbox{${A_1}$} }

\def\haone{\mbox{$h_{A_1}$}}

\def\dstarell{\mbox{$D^*-\ell$}}

\def\oomq{ {$1 \over m_Q$ } }

\def\eff{efficiency}
\def\mumc{\mu_{\rm mc}}

\def\Wtype{W_{\rm type}}
\def\ftype{f_{\rm type}}
\def\Ntype{N_{\rm type}}

\def\mut{\mu_{\rm t}}

\def\sigmadif{\sigma_{\rm dif}}

\def\rmc{r_{\rm mc}}
\def\MC{Monte Carlo}
\def\MCs{Monte Carlo simulation}

\def\nonum{\nonumber}

\def\calF{\mbox{$ {\cal F}$}}

\def\mthresh{m_{\rm thr}}
\def\fcomb{f_{\rm comb}}
\def\fother{f_{\rm otherPk}}
\def\fsignal{f_{\rm signal}}

\input pubboard/babarsymPRD

\def\spieff{\mbox{$\eps(p_{\spi})$}}

\def\cby        {\ensuremath{{\cos\theta_{BY}}}\xspace}
\def\dstst        {\ensuremath{{D^{**}}\xspace}}
\def\dst        {\ensuremath{{D^{*}}\xspace}}
\def\deltam        {\ensuremath{{\Delta m}}\xspace}
\def\ctl        {\ensuremath{{\cos\theta_\ell}}\xspace}
\def\ctv        {\ensuremath{{\cos\theta_V}}\xspace}
\def\angchi     {\ensuremath{{\chi}}\xspace}

\def\thetal        {\ensuremath{{\theta_\ell}}\xspace}
\def\thetav        {\ensuremath{{\theta_V}}\xspace}

\def\ctl        {\ensuremath{{\cos\theta_\ell}}\xspace}
\def\ctv        {\ensuremath{{\cos\theta_V}}\xspace}

\def\stl        {\ensuremath{{\sin\theta_\ell}}\xspace}
\def\stv        {\ensuremath{{\sin\theta_V}}\xspace}

\def\cali { {\cal{I}} }

\def\haonew{\mbox{$h_{A_1}(w)$} }

\newcommand{\BABARPubYear}    {05}
\newcommand{\BABARPubNumber}  {01}

\newcommand{\SLACPubNumber} {11156}
\def\figurebox#1#2#3{%
    \def\arg{#3}%
    \ifx\arg\empty
    {\hfill\vbox{\hsize#2\hrule\hbox to #2{\vrule\hfill\vbox to #1{\hsize#2\vfill}\vrule}\hrule}\hfill}%
    \else
    {\hfill\epsfbox{#3}\hfill}%
    \fi}
\def\rncversion{}
\def\rncdelete{}
\begin{document}


\preprint{\babar-PUB-\BABARPubYear/\BABARPubNumber} 
\preprint{SLAC-PUB-\SLACPubNumber} 

\begin{flushleft}
\babar-PUB-05/052 \\
SLAC-PUB-11672\\
hep-ex/0602023 \\ 
\end{flushleft}

\title{
{\large \bf
Measurements of the $B \rar \Dst$ Form Factors Using the Decay
$\Bbartodstarenu$ } }

%
\author{B.~Aubert}
\author{R.~Barate}
\author{D.~Boutigny}
\author{F.~Couderc}
\author{Y.~Karyotakis}
\author{J.~P.~Lees}
\author{V.~Poireau}
\author{V.~Tisserand}
\author{A.~Zghiche}
\affiliation{Laboratoire de Physique des Particules, F-74941 Annecy-le-Vieux, France }
\author{E.~Grauges}
\affiliation{IFAE, Universitat Autonoma de Barcelona, E-08193 Bellaterra, Barcelona, Spain }
% ...
\collaboration{The \babar\ Collaboration}
\noaffiliation

\newpage
\date{\today}

\begin{abstract}
We measure the dependence of $\bar B^0\rar D^{*+}e^-\overline{\nu}_e$
on the decay angles and momentum transfer.  The data sample consists
of $\sim 86 \times 10^6$ $B\bar B$-pairs accumulated on the
$\Upsilon(4{\rm S})$ resonance by the \babar\ detector at the
asymmetric $e^+ e^-$ collider PEP-II.  We specify the three form
factors by two ratios $R_1$ and $R_2$, and by a single parameter
$\rho^2$ characterizing the polynomial representing $h_{A_1}$, the
function which describes the momentum-transfer dependence of the form
factor $A_1$.  We determine $R_1$, $R_2$, and $\rhosq$ using an
unbinned maximum likelihood fit to the full decay distribution.  The
results are $R_1=1.396\pm 0.060\pm 0.035\pm 0.027$, $R_2=0.885\pm
0.040\pm 0.022\pm 0.013$, and $\rho^2=1.145\pm 0.059\pm 0.030\pm
0.035$.  The stated uncertainties are the statistical from the data,
statistical from the size of the Monte Carlo sample and the systematic
uncertainty, respectively.  In addition, based on this measurement, we
give an updated value for the CKM matrix element $|V_{cb}|$.

\end{abstract}

\vfill
\begin{center}

Submitted to Phys. Rev. D 

\end{center}

\maketitle

\section{Introduction}
\label{sec:Introduction}

The hadronic weak current of the exclusive semileptonic decay $B\rar
D^*\ell\nu$ with a light lepton $\ell$ = $e$ or $\mu$ can be described
by two axial form factors
\aone\ and \atwo, and one vector form factor $V$,
which are functions of the $B$-to-\Dstar\ momentum transfer squared,
$q^2$ (and though here we present results only for the $\bar B^0\rar
D^{*+}e^-\overline{\nu}_e$ decay, the results are expected to be the
same with the electron replaced by a muon, and thus apply to all
\Bbartodstarlnu\ decays where $\ell$ is a light charged lepton).
These form factors are usually characterized in terms of their ratio
\param s \rone\ and \rtwo, and a slope parameter \rhosq. These terms
will be defined precisely in Sec.~\ref{sec:Formalism}.  First
measurements of these three parameters were made by the CLEO
collaboration~\cite{CLEO}.  Improved measurements are important for a
significant reduction of the experimental error in the determination
of the Cabibbo-Kobayashi-Maskawa (CKM) matrix element $|V_{cb}|$.
They also provide for improved determinations of $|V_{ub}|$ from
inclusive and exclusive semileptonic $b\to u\ell\nu$ decays by
improving the accuracy with which the dominant $b\to c\ell\nu$
background to these decays is known.
                                                           
In heavy-quark effective field theory (HQET)
\cite{NeubertPhysReport,richmanburchat}, the three form factors are
related to each other through heavy quark symmetry (HQS). HQET allows
for three free parameters, which must be determined by experiment.
Deviations from the HQS relationships can be computed as corrections
to the theory. They can also be measured, and the \paramzn\ we adopt
is inspired by HQET, but allows for such deviations to be extracted
and compared against HQET predictions, as we will see in
Sec.~\ref{sec:Summary}.

The data used in this analysis were recorded by the \babar\ detector
at the \pep2\ storage ring, and correspond to $79 \ifb$ integrated on
the \upsfs\ resonance, yielding $ 86\times10^6$ $B\bar
B$-pairs. There are \mbox{$\sim 5 \times 10^6$} $\btodstarenu$ decays in this
sample, of which we have reconstructed 16,386 candidates for the decay
$\bar B^0\rar D^{*+}e^-\bar\nu_e$ using only the $\Dstar \rar
D^0\pi^+$ decay of the $D^*$ and the $\Dzero
\rar K^-\pi^+$ decay of the $D^0$ 
(the above also includes reconstruction of the corresponding charge
conjugate decay chain, and generally charge conjugation is implied
everywhere throughout this document).

We introduce here a novel method of extracting { form-factor}
parameters from the data. We use an unbinned maximum likelihood
method, and introduce approximations that allow us to correct for the
efficiency and resolution with the limited Monte Carlo (MC) data
sample available.  The impact of these approximations on our results
is studied in detail.

An important difference from the earlier analysis~\cite{CLEO} is that
in place of a linear
\paramzn\ of the $q^2$-dependence of the form factors
\haone, 
used for their main results,
we use a higher-order polynomial motivated by theory. 
The linear form was assumed in
similar analyses (in which only $|\vcb|$ and the slope \ff\ were
extracted, but not all three \ff s) by ARGUS~\cite{ARGUS}, operating
at the $\Upsilon$(4S), and by ALEPH~\cite{ALEPH1,ALEPH2} ,
DELPHI~\cite{DELPHI}, and OPAL~\cite{OPAL} at LEP.  The need for
higher-order terms has been indicated by recent data~\cite{babarVcb}.  Further,
this extension is also required by theoretical constraints
\cite{CLN}, \cite{GrinsteinLebed}, \cite{oliver}. The inclusion or exclusion of
higher-order terms leads to very different values for the parameter
$\rho^2$ (which is defined explicitly in Eq.~(\ref{eq:ffexpd}) ), as
will be discussed further in Sec.~\ref{sec:Summary}. In order to
compare with previous results, we also report the result of a fit
obtained with a linear \paramzn.

An outline of the paper is as follows: In Sec.~\ref{sec:Formalism}, {
we define the observables, form factors, decay amplitudes, and
parameters.}  Sec.~\ref{sec:babar} describes the relevant aspects of
the detector that are most significant for this analysis.  Event
reconstruction and selection is described in
Sec.~\ref{sec:recoAndSel}, the \MC\ simulation in Sec.~\ref{sec:mc},
and the analysis method in
Sec.~\ref{sec:Analysis}. Sec.~\ref{sec:Physics} describes fit results,
Sec.~\ref{sec:gof} the goodness-of-fit method and tests, and
Sec.~\ref{sec:systerrs} the estimation of systematic errors. In
Sec.~\ref{sec:Summary}, we summarize the results and compare them to
the previous measurements.  Appendix A covers technical details of the
pseudo-likelihood systematic error estimation method, and Appendix B
gives more detail on the determination of the levels of combinatoric
and peaking \bkgd s.

\hfill

\hfill

\hfill

\hfill

\section{Formalism}
\label{sec:Formalism}

\label{sec:wdescrip}

This section outlines the \rncdelete formalism and describes the
parameterization used for the form factors. More details can be found
in Refs. \cite{NeubertPhysReport,richmanburchat}.
 
The lowest order quark-level diagram for the decay \Bbartodstarlnu\ is shown in
Fig.~\ref{fig:quarkleveldiag}.  

\begin{figure}[ht]
\begin{center}
{\parbox{6cm}
{\resizebox{!}{3cm}{\includegraphics{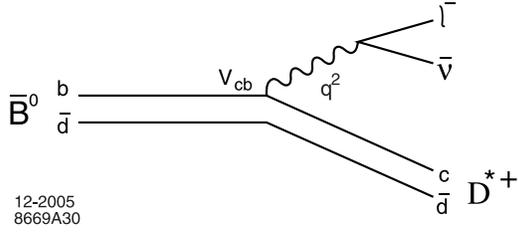}}
}}
\end{center}
\caption{Quark-level diagram showing the weak interaction vertices in the decay \Bbartodstarlnu.}
\label{fig:quarkleveldiag}
\end{figure}

\subsection{Kinematic Variables   }

A \Btodslnu\ decay is completely characterized by four variables,
three angles and $q^2$, the square of the momentum transfer from the
$B$ to the $D^*$ meson.

The momentum transfer is linearly related to another Lorentz invariant
variable, called $w$, by

\be
\label{eq:wdef}
\recow \equiv 
v_B \cdot \vdstr = { p_B \cdot p_{D^*} \over \mb \mdstar} =  {\mb^2 + \mdstar^2 -q^2 \over 2 \mb \mdstar},
\ee 
where $\mb$ and $\mdstar$ are the masses of the $B$ and the $D^*$
mesons, $p_B$ and $p_{D^*}$ are their four-momenta, and $v_B$ and
$v_{D^*}$ are their four-velocities. In the $B$ rest frame the
expression for $w$ reduces to the Lorentz boost to the \Dst, $\gamma_{D^*} =
E_{D^*}/M_{D^*} $.

The { ranges} of $w$ and $q^2$ are restricted by the kinematics of the
decay, with $q^2 \approx 0$ \rncdelete corresponding to
\be
w_{max}=\frac{\mb^2+\mdstar^2}{2\mb\mdstar}\approx1.504
\ee 
and $w_{min}=1$ corresponding to
\be
q_{max}^2=(\mb-\mdstar)^2\approx10.69 (\gevcc)^2.
\ee

In this analysis we only reconstruct the decay $D^{*+} \rar D^0
\pi^+$, where $D^0 \rar K^- \pi^+$ .  The angular variables, shown in
Fig.~\ref{fig:mesonleveldiag}, are

\begin{figure}[h]
\begin{center}
{\parbox {10cm}
{\resizebox{7cm}{!}{\includegraphics{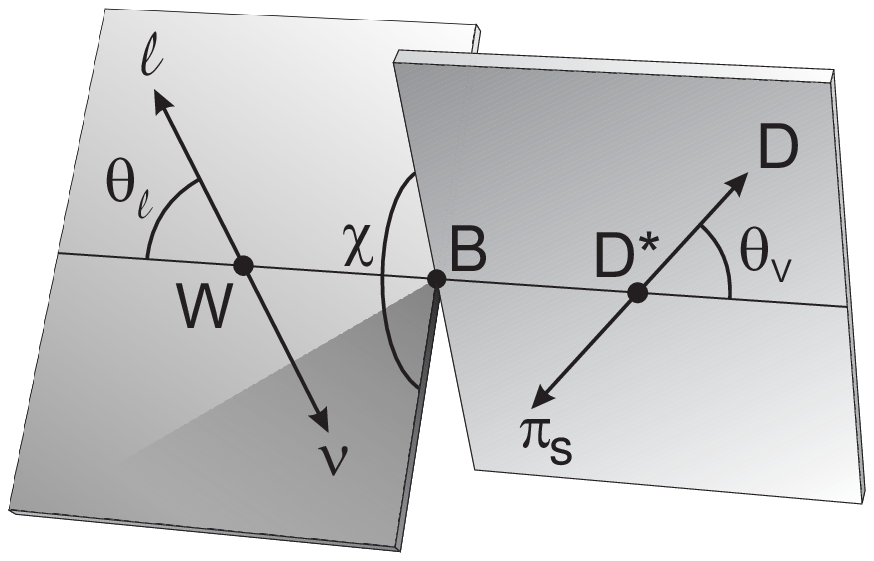}} }}
\end{center}
\caption{Kinematics of a \Btodslnu\ decay, mediated by a weak virtual intermediate vector boson $W$.  This diagram defines the three \kv\ angles \thetal, \thetav and \angchi.}
\label{fig:mesonleveldiag}
\end{figure}

\bi 

\item{ $\thetal$, the angle between the direction of the lepton ({i.e.,} for this analysis, the electron) 
in the virtual $W$ rest frame, and the direction of the virtual $W$ in
the $B$ rest frame }

\item{ $\thetav$, the angle between the direction of the 
$D$ in the $D^*$ rest frame, and the direction of the
$\dstar$ in the $B$ rest frame }

\item{$\angchi$,  the dihedral angle between the plane formed by the
$\dstar-D$ and the plane formed by the $W-\ell$ system. }
\ei

\subsection{ Four-Dimensional Decay Distribution}

The Lorentz structure of the $B\rar D^*\ell\nu$ decay amplitude can be
expressed in terms of three helicity amplitudes ($\ffH_+$, $\ffH_-$,
and $\ffH_0$), which correspond to the three polarization
states of the
\Dstar\ (two transverse and one longitudinal). 
For light leptons these amplitudes are expressed \cite{NeubertPhysReport,richmanburchat} in terms of the three form factors:

\bea
\label{eq:hel_ampp}
\ffH_{+}(w) & \equiv & -(\mb+M_{D^*}){A_{1}(w)}+ 2{ \momdstr \mb
\over \mb + M_{D^*}}{V(w)}, \nonumber
\\
\ffH_{-}(w) & \equiv & -(\mb+M_{D^*}){A_{1}(w)}-
      2{ \momdstr \mb \over \mb + M_{D^*}}{V(w)}, \nonumber
\\
\ffH_{0}(w)  & \equiv &- {\mb+M_{D^*} \over  M_{D^*} \sqrt{q^{2}}}  \left[ {A_{1}(w)}
        (M_{D^*}(w\mb-M_{D^*})) \right.   \nonumber\\
             & & \hspace{4cm} 
\hspace{-1cm} \left. -{4 \mb^2 \wp_{D^{*}}^2 \over (\mb + M_{D^*})^2 }{A_{2}(w)} ) \right] , \nonumber\\
\eea
where $\wp_{D^*} \equiv M_{D^*}\sqrt{w^2-1}$ is the magnitude of the momentum of the $D^*$ in the $B$ rest frame (for the $\tau$ lepton, which has non-negligible mass compared to the other particles in this process, a fourth helicity amplitude, $H_t$, would contribute~\cite{NeubertPhysReport} ).

The full differential decay rate in terms of the three helicity
amplitudes is~\cite{NeubertPhysReport, richmanburchat}

\bea
\label{eq:dsdo}
\label{eq:thlpdf}
{d\Gamma  \over dq^2 \; d\ctl \; d\ctv \; d\chi}  & = & 
{3G^{2}_{F}\vert V_{cb} \vert^{2}   \momdstr  q^2  \over 8(4\pi)^{4} M_B^2  }  {\cal B_{\rm D^* D}} \times 
 \nonumber\\
  & & \lb {H_{+}^2}(1-\ctl)^{2}\sin^{2}\theta_{V}+ \right.
\nonumber \\
&&   {H_{-}^2}(1+\ctl)^{2}\sin^{2}\theta_{V} +
\nonumber \\
 & & 4{H_{0}^2}\sin^{2}\theta_{\ell} \cos^{2}\theta_{V}  
\nonumber \\
 & &   \hspace{-0.6cm}    -2{H_{+}H_{-}}\sin^{2}\theta_{\ell} \sin^{2}\theta_{V}\cos2\chi  
\nonumber  \\
 & & \hspace{-2.6cm} -4{H_{+}H_{0}}\stl(1-\ctl)\stv\ctv\cos\chi   
\nonumber \\
 & &  \hspace{-2.6cm} \left. +4{H_{-}H_{0}}\stl(1+\ctl)\stv\ctv\cos\chi \rb , \nonumber
\\
\eea 
where ${\cal B_{\rm D^* D}} \equiv BF(\Dstarplus \rar \Dzero \piplus) \times BF(\Dzero \rar K^- \piplus)$ and where all three of the $H_i$ are functions of $w$. The
four-dimensional distribution of $w$, $\ctl$, $\ctv$, and $\chi$
described by Eq.(\ref{eq:thlpdf}) is the physical observable from
which we extract the form factors.

In this analysis we only deal with the shape and relative
normalization of the form factors, and consequently the overall
normalization of the rate is irrelevant.

\subsection{ Heavy Quark Symmetry Relations  }
\label{sec:hqsrships}

The functions $R_1(w)$ and $R_2(w)$ are defined in terms of the axial
and vector form factors by

\be
\label{eq:a2}
A_2(w) \equiv \frac{R_{ 2}(w)}{R^{*2}}\frac{2}{w+1}A_1(w),
\ee
and
\be
\label{eq:v}
V(w) \equiv \frac{R_{ 1}(w)}{R^{*2}}\frac{2}{w+1 } A_1(w),
\ee
where the constant $R^*$ is defined by
\be
R^*\equiv {   {2\sqrt{M_{B} M_{\Dstar}} }  \over {(M_B+M_{\Dstar})}   }.
\ee

Perfect HQS implies that $R_1(w)=R_2(w)=1$, {\it i.e.}, the form
factors $A_2$ and $V$ are identical for all values of $w$ {
and differ from $A_1$ only by a simple kinematic factor}.  Since HQS
is not exact, in general $R_1$ and $R_2$ { will} differ
from unity and exhibit some $w$-dependence as discussed in
Sec.~\ref{sec:ffpredics}.

It is conventional to introduce

\be
h_{A_1}(w) \equiv \frac 1{R^*}\frac2{w+1}{A_1}(w),
\ee
such that in the HQS limit, $h_{A_1}$ is the Isgur-Wise function $\xi(w)$ 
\cite{isgurwise} and
 $h_{A_1}(1)=\xi(1)=1$.

The function $h_{A_1}$ can be parameterized in a number of ways. One simple 
empirical expansion in $w-1$ is

\bea
\label{eq:ffexpd}
h_{A_1}(w) & = & h_{A_1}(1)  \times 
\\
& & \hspace{-1.5cm} \left(1-\rho^2(w-1)+\lambda (w-1)^2+ \kappa (w-1)^3+ ...\right), \nonumber
\eea

We will not fit with this general expansion as there are good
theoretical constraints (see Eq.~(\ref{eq:clnff})), which relate all
the higher-order \param s to the linear \rhosq\ \param.  Further,
convergence issues arise with attempts to fit a free functional form
with so many
\param s (see Sec.~\ref{sec:hofits}).

Corrections to HQS modify $ h_{A_1}(1) $ and thus lead to deviations
from the relation of $h_{A_1}(1)=1$.  In the baseline analysis we use
a parameterization of $h_{A_1}$ which conforms to requirements of
analyticity.  This was first developed and presented by Boyd,
Grinstein, and Lebed ~\cite{GrinsteinLebed}, though the form used here
employs a version of this parameterization proposed by Caprini,
Lellouch, and Neubert (CLN)~\cite{CLN}

\bea
\label{eq:clnff}
h_{A_1}(z) & = & h_{A_1}(1) \times
\\
& &  \hspace{-1.5cm} \left[ 1-8\rhosq z+(53\rhosq-15)z^2-(231\rhosq-91)z^3\right], \nonumber
\eea
where 
\be
z\equiv{{\sqrt{w+1}-\sqrt{2}} \over {\sqrt{w+1}+\sqrt{2}}}.
\ee

The maximum value of $z$ over the entire allowed range of $w$ is
$z_{max} = 0.04$.  In an expansion in $z(w)$ to $\order(w-1)$,
\rncdelete \rhosq\ is the coefficient of the linear term.  In this
case, \rhosq\ can be referred to simply as the slope
$\frac{dh_{A_1}}{dw}$ at $w=1$.  However, in higher-order expansions
of $z(w)$, the coefficient of $(w-1)^2$, called the ``curvature,''
will take on non-zero values, as will \coef s of $(w-1)^3$ and other
higher-order terms.  Thus, even though only the single parameter
$\rho^2$ is used, the functional form of Eq.(\ref{eq:clnff}) implies a
non-linear function for $\haone(w)$.

An alternative  parameterization motivated by HQS is 
\bea
\label{eq:oliver}
h_{A_1}(w) = \left(\frac{2}{w+1}\right)^{2\rho^2}. 
\eea
Le Yaouanc, Oliver and Raynal have demonstrated that this form obeys the bounds on the
derivatives they obtain in the HQS limit using a rigorous, general approach based on
Bjorken-like sum rules \cite{oliver}.

We do not use this form in our analysis, but note that in the $\rho^2$ range
of interest ($\sim 1-1.5$) it agrees with CLN to a few percent. Thus, $\rho^2$ 
obtained using this parameterization would be virtually identical to that
we obtain using Eq.(\ref{eq:clnff}).

\subsection{ { Form-Factor Ratios}  }
\label{sec:ffpredics}

As discussed in Sec.~\ref{sec:hqsrships}, for infinitely massive $b$
and $c$ quarks, HQS predicts $\rone=\rtwo=1$ exactly.  However, these
simple relations are broken for finite $b$ and $c$ quark masses,
because these ratios are modified by both perturbative
($\as$-dependent) and non-perturbative $\lqcdomx$ corrections, where
$m_x$ represents either the $b$ or $c$ quark mass.

Calculating higher-order loop corrections to the \FF s yields
expansions of the form~\cite{LigetiGrinstein}:

\bea
 {\rone}(w) & = & 1 + \lb \phi_{11} \as  + \phi_{12} \asq  \rb + \omega_1 \lqcdomx,  \\ 
 {\rtwo }(w) & = & 1  + \lb \phi_{21} \as  + \phi_{22} \asq    \rb + \omega_2 \lqcdomx.
\label{eq:correcs}
\eea

The coefficients $\phi_{ij}$ of the $\as$ terms are complicated expressions
of the \Dst boost $w$ which have been calculated perturbatively up to second order,
to an estimated accuracy of about one percent (see
\cite{LigetiGrinstein}).  Though they are not explicitly suppressed 
by a \oomq\ heavy quark mass corrections, they are all functions that approach
zero in the $w \rar 1$ HQS limit.  

The coefficients $\omega_i$ of the $\lqcdomx$ factors are called
``subleading Isgur-Wise functions.''
Subleading Isgur-Wise function correction terms are evaluated somewhat
differently in various models in the HQET framework, resulting in a
variety of predictions for $\rone(w)$ and $\rtwo(w)$.

Perturbative quantum chromodynamics (QCD) and a variety of other
theoretical tools have been employed to determine the behavior both at
and away from the $w=1$ endpoint. Close and Wambach used a simple
quark model \cite{CloseWambach} to find

\bea
  {\rone }( w) & = & 1.15-0.07( w-1), \label{eq:cwr1r2} \\
  {\rtwo }( w) & = & 0.91+0.04( w-1). 
\eea
Calculations with HQET  have produced a variety of results.  An early prediction by Neubert was~\cite{NeubertPhysReport}
\bea
  \rone(\recow) & = & 1.35-0.22(w-1)+0.09(w-1)^2,  \label{eq:neubr1r2} \\
 \rtwo(\recow) & = & 0.79+0.15(w-1)-0.04(w-1)^2.
\eea
More recently, CLN~\cite{CLN} used spectral functions, dispersion relations,
and HQS  to predict

\bea
 {\rone}( w) & = & 1.27-0.12( w-1)+0.05( w-1)^2,\\
 {\rtwo }( w) & = & 0.80+0.11( w-1)-0.06( w-1)^2.
\label{eq:CLNr1r2}
\eea
Ligeti and Grinstein \cite{LigetiGrinstein} using similar HQET
calculational tools find

\bea
 {\rone}( w) & = & 1.25-0.10( w-1), \label{eq:ligeti} \\
 {\rtwo }( w) & = & 0.81+0.09( w-1).
\eea

It can be seen that in all the predictions the \coef s of the
$(w-1)$ and $(w-1)^2$ terms are small; this is because \rone\ and
\rtwo\ are, by construction, ratios { that} are expected to
vary only slightly with $w$, whereas $\haone$ is not subject to such a
restriction.  For the predictions above, $R_1$ and $R_2$ vary by
$0.07$ or less over the full $w$ range.  For this reason, for our
baseline fit, we follow precedent in treating $R_1$ and $R_2$ as
constants, independent of $w$.  We will however examine deviations
from this baseline fit in Sec.~\ref{sec:Physics}.

\section{The \babar\ Detector}
\label{sec:babar}

The \babar\ detector is described elsewhere in
detail~\cite{ref:babar}.  This analysis uses four of the five
subdetectors of
\babar: the silicon vertex tracker, the drift chamber, a Cerenkov-light-based 
particle identification detector, and the electromagnetic
calorimeter. The analysis depends critically on the silicon vertex
tracker to reconstruct the low-momentum pions produced by the decay
$D^{*+}\rightarrow D^0\pi^+$, about two-thirds of which do not
traverse more than the first quarter of the drift chamber (as is
commonly done, we refer to these as ``slow'' pions, henceforth denoted
``$\spi$'').

\section{Reconstruction and Event Selection}
\label{sec:recoAndSel}

We \reco\ the electron track in the drift chamber and silicon
tracker and identify it using particle identification (PID)
information from { $dE/dx$ measured in the drift chamber}, photons
captured by the Cherenkov-light detector, and energy deposited in the
electromagnetic calorimeter.  The $D^ *$ is reconstructed through its
decay to a low momentum pion ($\pi_s$) and a $D^0$, and the \Dzero\
through its decay to $K\pi$.  The hadrons are selected by similar PID
information to that used for the electron identification.

We then choose final cuts that select $B\rar D^* e \nu$ decay
candidates, and from the four-momenta of the observed particles we
{determine} the kinematic variables $w$, $\ctl$, $\ctv$, and $\chi$.

\subsection{\Bkgd s and Event Selection}
\label{sec:bkgdcats}
 
\sssec{\Bkgd\ Categories}
\label{sec:bkgdcatsdetail}
{We address each distinct source of background with
appropriate cuts (which are further described below in Sec.~\ref{sec:evtselcuts}):}

\begin{enumerate}

\item {Combinatorial background: events in which the reconstructed $D^*$ 
candidates that were not
originally actual
\dst\ mesons. These events do not contribute to the peak in the
$\Delta m=m_{K\pipi}-m_{K\pi}$ distribution.  Cutting on $\Delta m$ provides
discrimination against this background.}
\item Peaking
background, for which the $D^*$ decay has been correctly reconstructed
and which contributes to the peak in the $\Delta
m$ distribution. This category is further broken
into two main sub-categories: 

\begin{enumerate} 

\item  \Dstst\ \bkgd:   Events 
where a true $D^*$ is combined with a electron from the same $B^0$ or
$B^-$ parent, but an extra pion in the decay has been missed.  This is
primarily {feed-down} from P-wave $D$ meson decays, but also includes
non-resonant $B \rar D^* e \nu X$ decays, where $X$ is
\mbox{$n \pi$} (\mbox{$n \geq 1$}).  As shorthand we call this ``\Dstst\ \bkgd'' after its dominant component.  Note though that the four P-wave $D$ meson states are usually referred to by the shorthand term ``\Dstst\ modes'' in the literature, and thus our redefinition for the purposes of this paper of ``\Dstst\ \bkgd'' differs slightly from the standard terminology.

\item {Other events with a true $D^*$:}

\begin{itemize}

\item
\underline{Fake electron}: Events in which a real $D^*$ is combined with a hadron instead 
of an electron.  This background is minimized by requiring electrons
to pass the most stringent electron-identification criteria.
\item
\underline{Cascades}: Events with a decay chain of the form $\Bzerobar$ or $B^- \rightarrow
\Dst X$, $X \rightarrow e Y$ (X is {\it e.g.}, another \Dstar\ or
\Dzero, Y is anything), so that the observed electron is not primary 
({\it i.e.}, not directly from a $B$ decay).  Secondary electrons have a
softer momentum spectrum than primary electrons so that a minimum electron
momentum cut is effective against this \bkgd\ source.
\item
\underline{Uncorrelated electron}: Events where a $D^*$ is combined with a electron from 
the other $B$.  Thus, one side has a $\Bzerobar$ or $B^- \rightarrow
D^*X$, and the other has $\Bzero$ or $B^+ \rightarrow Y e$.
A cut on $\cosby$, the cosine of the angle between the $B$ and
the $\Dstar-e$ combination, is effective against this background.

\item
\underline{Continuum}:  $\epem \rar\ c\overline{c}$ events for which one $c$ quark 
forms a $D^*$ while the other hadronizes into a state which decays
semileptonically to create an electron ($c\overline{c} \rightarrow
D^*e X$).  The other continuum backgrounds ($\epem \rar u\ubar,d\dbar,s\sbar$) are
negligible and almost none of them pass the final
cuts.  Cuts on the $D^*$ momentum and
event topology are effective at suppressing this background.
\end{itemize}
\end{enumerate}
\end{enumerate}

\sssec{Event Selection Cuts}
\label{sec:evtselcuts}

For event selection we use the procedure developed for {the \babar\ $V_{cb}$
analysis~\cite{babarVcb}. The most salient cuts are as follows:}

\begin{itemize}

\item{
The momentum of the electron in the center-of-mass (C.M.) frame, which
is denoted throughout this paper as ``$p^*_\ell$'', is required to be
larger than $1.2
\gevc$. This criterion selects $B$ semileptonic decays and suppresses
continuum ($\eplus\eminus \rar c\bar c$) and cascade ($B \rar D \rar e
$) backgrounds.  }

\item{

The slow pion from the $D^*$ decay must have a transverse momentum
$p_t$ greater than $50 \mevc$. This rejects the mostly fake tracks
found below this cut. The efficiency for finding true pion tracks
below $50 \mevc$ is small as the majority of the pions stop before
leaving enough hits in the vertex detector to be reconstructed.

\item{
The $\chi^2$ probability of the fit of the $D^* e $ vertex{,
including the beam-spot contraint,} must be greater than $1\%$.  This
suppresses inclusion of particles from the other $B$, {\it e.g.}, those
tracks in uncorrelated electron background events.

\item {To further suppress continuum background, 
we select only candidates with $|\cos\theta_{\rm thrust}|<0.85$, where
$\theta_{\rm thrust}$ is the angle between the thrust axis of the
$D^* e$ candidate and the thrust axis of the rest of the event.
}

\item{ The cosine of the angle $\theta_{BY}$ between the direction of the $B$ and the
direction of the \dstarell\ system can be computed from the kinematics
of the \btodstarlnu\ decay (see Sec.~\ref{sec:wdetermination}).
Candidates with $\cosby$ between $-10$ and $+5$ have been used to
estimate background.  We include only events that have $|\cosby| \leq
1.2$ in the final sample. The cut is set beyond the physical limits at
$\pm 1$ to allow for spillover due to resolution.}}

\item{ The final selection is based on $\Delta m \equiv m_{K\pi\pi_s}-m_{K\pi}$ (the difference between the \recod\ \Dstar\ and \Dzero\ masses). We 
require $0.143\leq\Delta m\leq 0.148$ {\bf \gevcc}.}}

\end{itemize}

\begin{figure}[htp]
\begin{center}
{\parbox{7cm}
  {\resizebox{!}{6cm}{\includegraphics{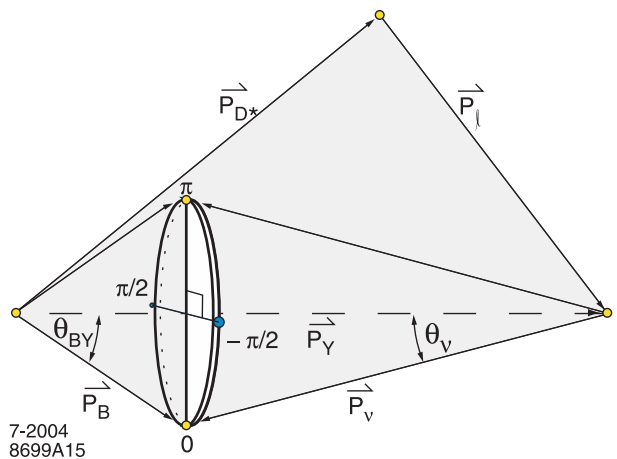}}
}}
\end{center}
\caption{\label{fig:cosbyfig}
 Reconstructed particle directions and $\cos\theta_{BY}$
{in the CM frame.  The momentum vectors $\vec{\bf P}_{D^*}$
and $\vec{\bf P}_\ell$ are measured.  Their sum is $\vec{\bf P}_Y$.  The magnitude of
the $B$ momentum, $\wp_B \equiv |\vec{\bf P}_B|$, and the angle $\theta_{BY}$ the $B$ makes with $\vec{\bf P}_Y$ are known, but the azimuthal orientation $\phi_{BY}$ of $\vec{\bf P}_B$ around $\vec{\bf P}_Y$ is not known.}
The points at $\phi_{BY}=0, \pi$
are in the $D^*-\ell$ plane.
}
\end{figure}

\subsection{ Determination of Kinematic Variable \boldmath{$w$} }
\label{sec:wdetermination}

{Lacking a measurement of the neutrino momentum}, we do not have sufficient information to fully
reconstruct the kinematic variables $w$, $\ctl$, $\cos\theta_V$, and
$\chi$. However, using energy-momentum conservation and assuming that
the missing particle is a massless neutrino, we have

\be
\label{eq:mnu}
0=m^2_\nu=M_B^2+M_Y^2-2E_BE_Y+2\wp_B \wp_Y \cos\theta_{BY},
\ee
where $p_Y \equiv p_{D^*}+p_\ell$ is the four momentum of the combined $D^*$
and electron, $M_Y^2=p_Y^2$ is the mass squared and $\wp_Y$ is the
magnitude of the $Y$ three-momentum.  The $B$ meson energy $E_B$ and
three-momentum magnitude $\wp_B$ {are known} from the energies of the
colliding beam particles, so we can solve for $\cos\theta_{BY}$:

\be
\label{eq:cosby}
\cos\theta_{BY}=-\frac{M_B^2+M_Y^2-2E_BE_Y}{2 \wp_B \wp_Y }.
\ee

Thus we can determine the angle between the $B$ and the direction
($\hat Y=\vec p_Y/\wp_Y$, where $\vec p_Y$ is the three-vector
momentum of the $Y$) of the \dstarell\ system, but we do not know the
azimuthal angle $\phi_{BY}$. This is illustrated in
Fig.~\ref{fig:cosbyfig} where it can be seen that the direction of the
$B$ must lie on the cone centered on $\hat Y$ with the opening angle
$\theta_{BY}$.

For each possible $\phi_{BY}$ we can compute the kinematic variables
$w$, $\ctl$, $\ctv$, and $\chi$. Since the angle $\phi_{BY}$ is not
measured, we average over four points: two in the $D^*$-electron
plane corresponding to the azimuthal angles $\phi_{BY}=0$ and $\pi$
and two points out of the plane corresponding to the angles
$\pm\pi/2$. Further, since $B\bar B$ production follows a
$\sin^2\theta_B$ distribution in the angle between the $B$ direction
and the beam collision axis in the CM (\upsfs) frame, we weight the
\kv s evaluated at each point by  $\sin^2\theta_B$.

\begin{figure}[hbp]
\begin{center}
{\parbox{8cm}
  {\resizebox{!}{8cm}{\includegraphics{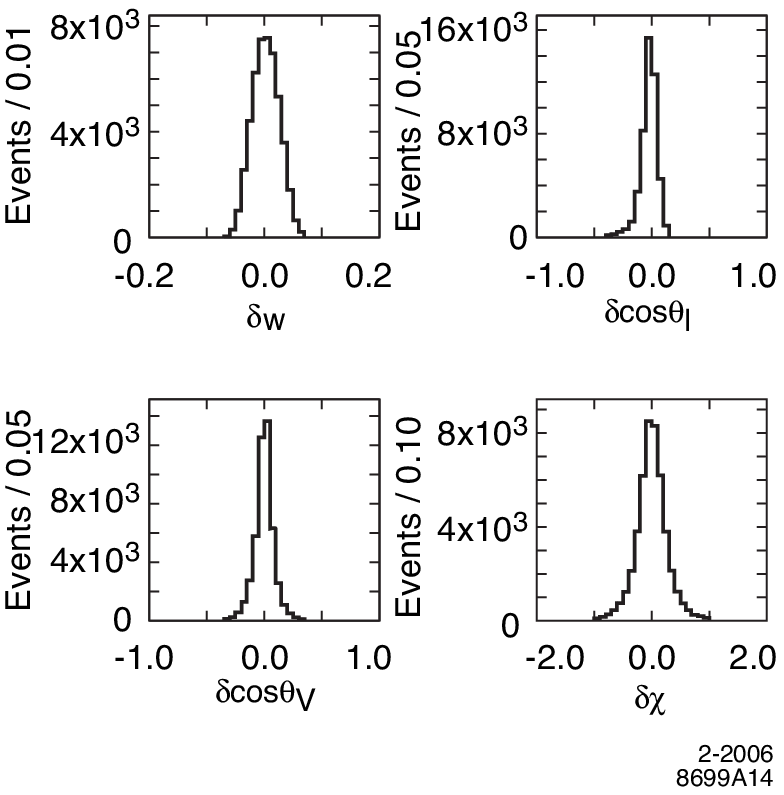}}
}}
\end{center}
\caption{\label{fig:kinvarres}
\MC\ assessment of the experimental resolution for the variables $w$,
$\ctl$, $\ctv$, and $\chi$. For each variable the difference
between reconstructed and generated values is shown.  The \resln s are
small compared to the kinematic ranges of the variables as shown
in Fig.~\ref{fig:kinvar-plot}. }
\end{figure}

Fig.~\ref{fig:kinvarres} illustrates the resolution achieved by this
technique.  The core widths for each \resln\ \dist\
are small compared to the full range of each \kv.  The
resolution is dominated by the average over the $B$ direction;
detector resolution makes a relatively minor contribution. The
low-side tail on $\ctl$ can be attributed to final state radiation.

The resolutions of the four kinematic variables are highly
correlated. Thus, we rely on Monte Carlo simulation to account for
resolution effects.  

\begin{figure}[ht]
\begin{center}
{\parbox{8cm}
  {\resizebox{!}{8cm}{\includegraphics{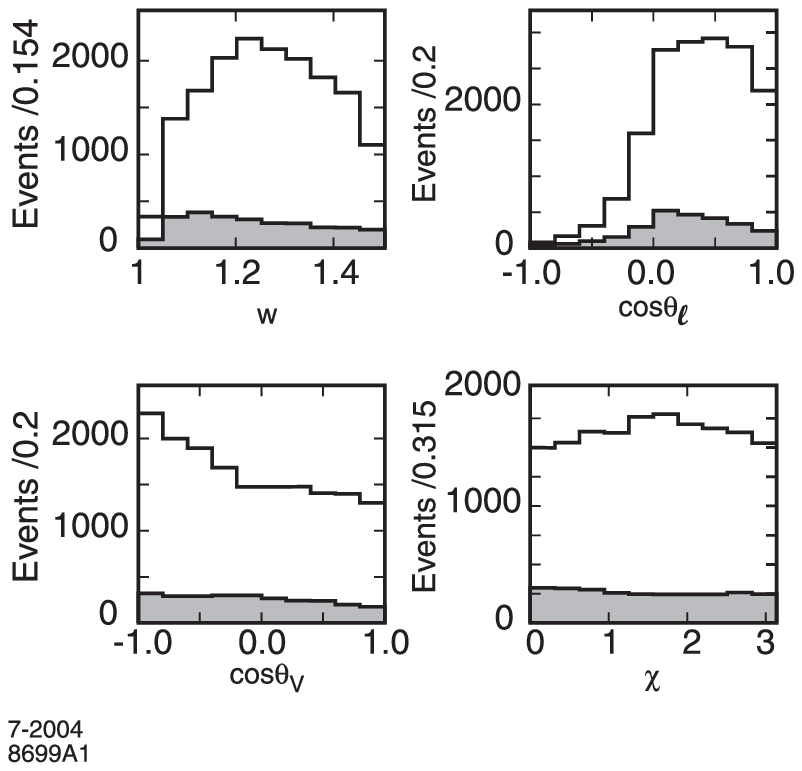}}
}}
\end{center}
\caption{\label{fig:kinvar-plot}
One-\diml\ projections from \MCs\ of kinematic variables $w$, $\ctl$,
$\ctv$, and $\chi$ for selected events.  The blank region represents
signal, and the shaded regions are estimated backgrounds; the \histo s
are a sum of both the signal and the estimated \bkgd\ beneath them.}
\end{figure}

The distributions of the reconstructed kinematic variables $w$,
$\ctl$, $\ctv$, and $\chi$ from \MCs\ are displayed in
Fig.~\ref{fig:kinvar-plot}. The shaded region is the distribution of
the background as estimated from the \MC\ simulation using the method
described in Sec.~\ref{sec:Analysis} below.  The \bkgd\ contributions
to the \dist s are much smaller than the signal contribution (on the
order of 10-15\%).

\section{Simulation}
\label{sec:mc}

This analysis is dependent on Monte Carlo (MC) simulation to model the
efficiency and the background distributions.  The degree to which the
simulation reproduces both the detector response and the underlying
physics processes largely determines the systematic errors.

The response of the BABAR detector is modeled using a GEANT4-based
simulation.  ~\cite{babarsim}. The simulation has been extensively
validated by comparison with large data control samples (such as
slow pions from generic $\Dstar \rar \Dzero \pi$ decays for the
slow pion helicity studies).  Event generation and particle decay
are modeled using the package EvtGen~\cite{evtgen}.

\begin{figure}[htp]
\begin{center}
{\parbox{8cm}
  {\resizebox{!}{8cm}{\includegraphics{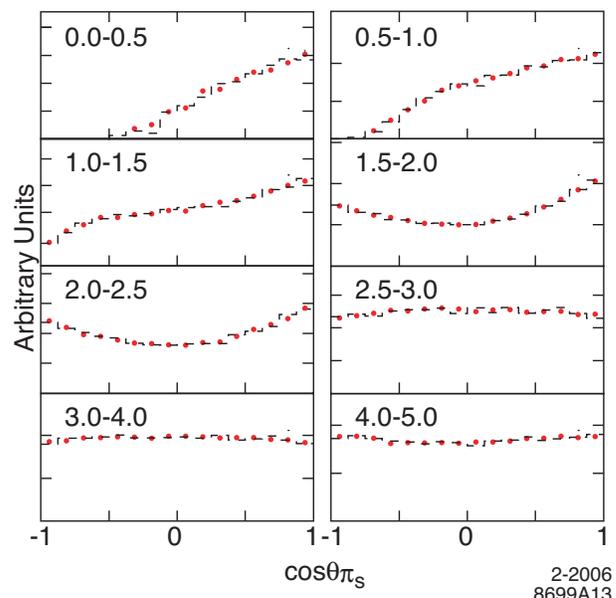}}
}}
\end{center}
\caption{ \label{fig:dsthel}
Events vs. the cosine of the slow pion angle relative to the
line-of-flight of the $D^*$, shown for bins of CM momentum of the
$D^*$.  The momentum ranges for each plot, in GeV/c, are indicated in
the upper left-hand corners.  Data (points) and MC simulation
(histogram) are shown for slow pions from the decay of inclusively
produced $D^*$ mesons.  The vertical scale is arbitrary, but
corresponds to the number of background-subtracted reconstructed $D^*$
candidates in a given bin. Only the differences between data and MC
points are relevant for our purposes.  The errors are comparable to the dots
representing the data.  The simulation has been weighted by restricted
functions to fit to the data as described in the text.}
 
\end{figure}

\ssec{Signal}
To simulate the signal we use Eq.~(\ref{eq:dsdo}) for the distribution
of the decay products.  The MC samples are generated with the default
parameters $R_1=1.180$, $R_2=0.720$ and $\rhosq=0.920$ \cite{Ryd}. The
MC generator uses a simple linear expansion (Eq.~(\ref{eq:ffexpd})
taken to order $(w-1)$), so we must reweight the MC events to model a
more complex behavior such as that given by Eq.~(\ref{eq:clnff}).

\ssec{Detection of Slow Pions}
\label{sec:slowpi}
Of particular importance to this analysis is the modeling of the
efficiency for detecting low-momentum pions. This task requires
detailed simulation work since low momentum pions are lost through the
interplay of acceptance, decay-in-flight, and stopping and scattering
in the beam pipe and vertex detector.  

To test this modeling we examine a large control sample of inclusively 
produced $D^{*+}$ mesons. Fig.~\ref{fig:dsthel} shows the distribution 
of cosine of the helicity
angle for the decay $D^{*+}\rightarrow D^0\pi^+$ reconstructed in 
both the data and the MC sample in bins of
$D^{*+}$ center-of-mass momentum. The helicity is defined so that slower pions 
correspond to backward helicity angle (and is $\cong -\ctv$).

The MC distributions have been fit to the
data using a simple weighting factor of the form
\be
\label{eq:fcorr}
f_{\rm corr}=N\left( 1+\xi_j \cos\theta_{\pi_s}+ \zeta_j \cos^2\theta_{\pi_s} \right),
\ee
where the normalization $N$ and the parameters $\xi_j$ and $\zeta_j$
have been obtained for each momentum bin ${ j}$.  The three lowest
momentum bins are most relevant to $B \rar D^*\ell\nu$ decays.  The
agreement between the weighted { simulation} and the data is
excellent.  Note the weighting factor Eq.(\ref{eq:fcorr}) is not
applied to correct the signal. Rather the weight coefficients obtained
by fitting this control sample measure the MC-data differences in each
momentum bin.

The linear terms, $\xi_j$, arise from inadequacies in the simulation
of the detector response.  The linear terms for the first three bins
are small: $0.037\pm 0.160$, $-0.023\pm 0.024$ and $-0.016 \pm 0.009$,
indicating good agreement between the data and
\MCs\ efficiencies in the complicated arena of slow pion modeling. Note that 
critical details of the distributions that are unaffected by the
weighting scheme (such as the evident thresholds in the low momentum
bins) are reproduced correctly by the MC.

The quadratic terms arise mostly from incorrect simulation of the
polarization of the $D^*$-mesons. The values of $\zeta$ are large
($\sim 0.5$ and $0.7$) in the momentum bins for $0.5-1.0$ GeV/$c$ and
$1.0-1.5$ GeV/$c$, where we expect $D^*$-mesons to be highly polarized
products of few-body $B$-meson decays. These terms do not indicate any
problem with detector simulation, and only affect a tiny fraction of
the events in our final sample (the fake electron background).

Further, as the MC and data agree well for the linear terms, there is
no need to make corrections to account for differences between
simulation and data.  This is further indicated by the lower right
plot of $p_{t,\spi}$ shown in Fig.~\ref{fig:uncutkinvar}, which indicates
excellent agreement between the slow pion transverse momentum \dist\
after the final fit, but with no linear weighting terms applied to the
slow pion momentum.

As the extraction of \FF s (especially \rhosq) is highly sensitive to
the slow pion simulation we do add a contribution to the
\systerr\ to account for the possible residual differences between the
data and the MC simulation.  We describe our method for doing this in
Sec.~\ref{sec:spierr}.

\ssec{Final State Radiation}

The program PHOTOS~\cite{Was} is used to model the effects of final
state radiation (FSR). PHOTOS uses quantum electrodynamics (QED) to
second order in $\alpha_{em}$ (up to two FSR photons can be produced)
and is known to provide a quite accurate simulation of the FSR effects
\cite{WasII}.

\ssec{Other Semileptonic Decays}
A major source of background is other semileptonic $B$ decays. Only
the branching fractions for the decay modes $D\ell\nu$, the signal
mode $D^*\ell\nu$, and that for the mode $B \rar D_1\ell\nu$ have been
measured~\cite{pdg2004} ({\it i.e.}, only one of the four P-wave $D$ meson
states).  The other branching fractions (including the three other
P-wave $D$ states and all non-resonant decays) used in the MC simulation
are based on models with large theoretical uncertainties.  We use fits
to the data to constrain these branching fractions as shown in
Sec.~\ref{sec:systerrs}, to minimize their contribution to the
systematic background error.

\section{Analysis Method}
\label{sec:Analysis}
\subsection{Fitting}
\label{sec:fitting}

Our approach to extracting the form factors is to perform an unbinned
maximum likelihood fit to the full four-dimensional distribution
function (PDF) specified by Eq.~(\ref{eq:dsdo}).  We parameterize the
form factors in terms of the parameters $R_1$, $R_2$ and $\rhosq$ as
described in Sec.~\ref{sec:Formalism}.  { The $\rhosq$ dependence is
specified by Eq.~(\ref{eq:clnff})}.  Since theoretical predictions
show only a mild dependence of $R_1$ and $R_2$ on $w$ (see
Eqs.~(\ref{eq:cwr1r2}-\ref{eq:neubr1r2}), we first perform the fit
treating them as constants over the entire range of $w$.  We later
show (in Sec.~\ref{sec:Physics}) how the results vary when the
$w$-dependencies suggested by the
Eqs.~(\ref{eq:cwr1r2}-\ref{eq:neubr1r2} are imposed.

In addition to assuming its form, we must account for the effects of
resolution and efficiency on the measured distribution.  We adopt
approximations that enable us to carry out the maximum likelihood fit
to the \FF\ \param s efficiently with minimal loss in precision.

\sssec{Resolution-Efficiency Correction Method}
\label{sec:recmethod}

To account for the efficiency and resolution effects we adapt the
approach first employed in the angular analysis of the decay
$B\rightarrow J/\psi K^*$\cite{psikstar} (for more details of the
technique see also~\cite{Gill}). The full PDF (${\cal F}$) including
resolution and efficiency is given in terms of the theoretical PDF ($F$) of Eq.~(\ref{eq:thlpdf}) by

\be
\label{eq:respdf}
{\cal F}(\tilde x;\mu)=\int{dx \, \varepsilon(x) \, G(\tilde x; x) \, F(x;\mu)},
\ee
where $x$ represents the true variables ($w$, $\ctl$, $\ctv$, and
$\chi$), $\tilde x$ are the observed values of the variables and
$\mu$ represents the parameters ($R_1$, $R_2$ and $\rhosq$) that
determine the form factors.  The efficiency $\varepsilon(x)$ is the
fraction of events with parameters $x$ that are
detected and $G(\tilde x; x)$ is the probability density that an event with true
parameters $x$ is reconstructed with parameters $\tilde x$.

The logarithm of the likelihood $L$ that we need to maximize is given by

\bea
\label{eq:likelihood}
\ln L=\sum_i{\ln\left( \frac{{\cal F}(\tilde x_i;\mu)}{\cali(\mu)}\right)}
\\
=\sum_i{ \ln {\cal F}(\tilde x_i;\mu)}-N_{\rm { data}}\times \ln \cali(\mu),
\nonumber
\eea
where the integral 
\be
\cali(\mu)\equiv\int{d\tilde x{\cal F}(\tilde x;\mu)}
\ee
is required to normalize the likelihood in
the presence of imperfect acceptance.  The sum is over our
data sample of $N_{\rm { data}}$ events.

{ While the distribution ${\cal F}(\tilde x,\mu)$ can be simulated by
Monte Carlo, it is not practical to use it directly as the function to
be varied in the maximum likelihood analysis.  Instead, we will
introduce a method in which we can use \MCs\ data without any need to
extract a detailed model of the efficiency and resolution function.

We now try the approximation
\be
{\cal F}(\tilde x;\mu) \cong f(\tilde x;\mu)\equiv{\cal F}(\tilde x;\mu_{\rm mc})\frac{F(\tilde x;\mu)}{F(\tilde x;\mu_{\rm mc})}.
\label{eq:approx}
\ee
where $\mu_{\rm mc}$ is the parameter set used for generation of the
Monte Carlo sample.  If we had used the true values of the parameters
$\mu_{\rm t}$ in place of $\mu_{\rm mc}$, then the maximum likelihood
method would converge to the true values since by inspection the trial
function $f(\tilde x;\mu)$ would become the true distribution 
${\cal F}(\tilde x;\mu)$.
Of course the true values are not known. The use of $\mu_{\rm mc}$
introduces a bias proportional to the difference between the Monte
Carlo parameter values and the true parameter values, which can be
calculated explicitly in the limit of high statistics.

When the approximation  $f(\tilde x;\mu)$ of Eq.~(\ref{eq:approx}) 
is substituted into the expression for 
the log-likelihood Eq.~(\ref{eq:likelihood}), it yields

\bea
\label{eq:llapprox}
\ln L & = & \sum_i \ln F(\tilde x_i;\mu)-\sum_i \ln F(\tilde x_i;\mu_{\rm mc})
\\
&&+ \sum_i \ln {\cal F}(\tilde x_i;\mu_{\rm mc}) -N_{\rm { data}}\ln \hat\cali(\mu,\mu_{\rm mc}),
\nonumber
\eea
where $\hat\cali(\mu,\mu_{\rm mc})$ is the  integral of the approximation of Eq.(\ref{eq:approx}).}

Since terms that are independent of the fit parameters (constant
terms) do not affect the point at which the maximum will be found, all
the sums that depend only on $\mu_{\rm mc}$ can be dropped ({\it
i.e.}, the central two terms of the total sum). The $\mu$-dependent
piece has been factored from these constant terms. We are left with a
likelihood \funct\ that depends only on the theoretical PDF $F$ and on
the integral over the resolution and efficiency functions.

{ The resolution and efficiency occur only through  the normalization integral
$\hat\cali(\mu,\mu_{\rm mc})$}.

Using the technique of Monte Carlo integration to
evaluate the integral $\hat\cali(\mu,\mu_{mc})$ gives

\be
\label{apprxintegral}
\hat\cali(\mu,\mu_{\rm mc}) = \int{d\tilde x{\cal F}(\tilde x;\mu_{\rm mc})\times\frac{F(\tilde x;\mu)}{F(\tilde x;\mu_{\rm mc})}} 
\ee
\be
	\approx\frac{1}{N_{\rm gen}}\sum_i{\frac{F(\tilde x_i;\mu)}{F(\tilde x_i;\mu_{\rm mc})}}. \nonumber
\ee
The MC simulation generates events in proportion to ${\cal F}(\tilde
x;\mu_{\rm mc})$ so the sum over \MC\ events approximates the desired
integral.  The small error introduced by the Monte Carlo evaluation of
the normalization integral is determined in Appendix A.

In Appendix A, we demonstrate that the bias introduced by the use of
the approximate form Eq.(\ref{eq:approx}) is given by

\be
\label{eq:bias}
(\mu -\mu_{\rm t})_a=\sum_b(\mumc-\mu_{\rm t})_b M_{ba}
\ee
where
\def\expect#1{{\langle #1\rangle}}
\be
\label{eq:bias}
M_{ba}=J_{bc}E_{ca},
\ee
and
\bea
J_{bc}& \equiv &
{\left<{ \frac{\partial \ln\frac{\cal F}F}{\partial \mu_b}\frac{\partial\ln F }{\partial \mu_c}}\right>
-\left<{\frac{\partial \ln\frac{\cal F}F}{\partial \mu_b}}\right>
\left<{\frac{\partial \ln F}{\partial \mu_c}}\right>}\nonumber\\
 E^{-1}_{ac}& \equiv &
{\left<{\frac{\partial \ln F}{\partial \mu_a}{\frac{\partial \ln F}{\partial \mu_c}}}\right>
-\left<{\frac{\partial \ln F}{\partial \mu_a}}\right>
\left<{\frac{\partial \ln F}{\partial \mu_c}}\right>}
\eea
and where averages are defined by
\be
\expect A \equiv \frac{\int dx {\cal F}(x,\mu_{\rm t})A(x)}{\int dx{\cal F}(x,\mu_{\rm t}) }.
\ee

The bias vanishes if the parameters used in the Monte Carlo coincide
with the true values.  Moreover, the bias vanishes insofar as the
ratio ${\cal F}/F$ of the smeared distribution to the unsmeared
distribution is independent of the parameters $\mu$.  Since ${\cal F}$
and $F$ probe nearby regions, the derivative of their ratio will be
small, so the coefficient $\alpha$ in Eq.~(\ref{eq:bias}) should be
much less than one. Numerical evaluation yields values of $\sim 0.15$
for $R_1$ and $R_2$ and $\sim 0.03$ for $\rho^2$.  If $\mumc-\mut$ is
comparable to the error, the residual ``bias'' should be small, and in
practice is found to be so.

To achieve $\mumc\sim\mut$ we reweight the MC used in the efficiency
integral computation to the fitted values of the parameters and
iterate until the fit values converge.  At this point
$\mumc=\mu_{fitted}$ and should deviate from the truth by an amount
comparable to the error. In effect the residual deviation acts like a
small increase ($\sim 2\%$) in the statistical uncertainty.  The
iteration method works well and converges quickly, as discussed
further in Sec.~\ref{sec:iterativeconv}.

\subsubsec{Speeding Computation of Normalization Integral Through Moments Factorization}

Because the normalization integral $\hat\cali (\mu,\mu_{\rm mc})$
depends explicitly on $\mu$ it must be recomputed for every iteration
of the procedure that maximizes the log-likelihood.  To do this by
Monte Carlo integration over the full decay phase space would be
prohibitively slow.  

For the signal distribution, 
this integration can be avoided.
Since the signal PDF can be written in the following form,

\be
\label{factorizedpdf}
F(\tilde x;\mu)=\sum_\alpha{A_\alpha(\mu)\times\Xi_\alpha(x)},
\ee
{\it i.e.}, as sum over a product of terms depending only on the fit
parameters and terms depending only on the kinematic variables, we can
define moments $M_\alpha$ by

\be
\label{moments}
M_\alpha=\frac{1}{N_{gen}}\sum_i {\frac{\Xi_\alpha(\tilde x_i)}{F(\tilde x_;\mu_{\rm mc})}},
\ee
where the sum is over reconstructed MC events, {\it i.e.}, the same sum that
defines { $\hat\cali(\mu,\mu_{\rm mc})$} in Eq.~(\ref{apprxintegral}).  
This allows us to
write { $\hat\cali(\mu,\mu_{\rm mc})$} as a sum over moments:

\be
\label{momentsum}
{ \hat\cali(\mu,\mu_{\rm mc})}=\sum_\alpha{A_\alpha(\mu)\times M_\alpha}.
\ee 
The moments can be computed once before fitting and then taken as
input to the fit. Thus, in the fit, the time-consuming sum over
weighted events is replaced with the sum over moments.  Taking the
expansion of $h_{A_1}$ to order $(w-1)^3$ or $z^3$, we have 42 moments
to compute and sum.

\sssec{{ Background Subtraction Through Pseudo-Likelihood}}

To handle the background, we would ordinarily 
add a PDF $B(\tilde x)$ that models the background. The PDF
${\cal F}$ would be replaced with
\be
f{\cal F}(\tilde x;\mu)+(1-f)B(\tilde x),
\ee
where $f$ is the signal fraction.  However, since we do not have a
form for the background distribution before acceptance, we cannot
achieve the factorization of the parameter dependence from the
efficiency and resolution functions that leads to {
Eq.~(\ref{momentsum})}.  To avoid this problem we use the technique of
subtracting Monte Carlo events representing the background directly in
our likelihood sum rather than adding it to our PDF. We replace our
log-likelihood function with the following
`pseudo-likelihood'~\cite{BbrPseudoL}

\bea
\label{pseudoll}
\ln \Lambda &=& \sum_{i \in {\rm data} } \ln F(x^{(i)}_{\rm data};\mu)- 
\\
&& \sum_{j \in {\rm MC}} W^{(j)}_{\rm bkgd}\ln F(x^{(j)}_{\rm bkgd};\mu)- N_{{ \rm signal}}\ln \cali(\mu),
\nonum
\eea
where the first sum is over the data and the second is over a Monte
Carlo sample representative of the background. The weights
$W^{(i)}_{\rm bkgd}$ account for any difference between the background
in the data and in the Monte Carlo.  They are computed in the manner
indicated by Eqs.~(\ref{eq:bkwt}) and (\ref{eq:correctwt}) in
Sec. \ref{sec:back} below. { The coefficient of the normalization
integral is $N_{\rm signal}=N_{\rm data}-N_{\rm bkgd}$, where $N_{\rm bkgd}$ is equal
to the number of events subtracted, accounting for the weights.  It is
easy to show that if the Monte Carlo simulation of the background is
accurate, the procedure is unbiased.  The statistical errors can be
computed by a modification of the procedure used for an ordinary
log-likelihood analysis.}

\subsection{Methodological Error Contributions}

The approximations outlined above 
provide a fast and easily-implemented fitting procedure that uses the
MC sample without any need to extract a detailed model of the
efficiency and resolution functions.  However, these advantages do not
come without a cost: we must account for the uncertainty introduced by
the resolution-efficiency and pseudo-log-likelihood procedures.

There are three other contributions to the error that are not
accounted for in the fit. Two of these are Monte Carlo statistical in
nature: the error from the { pseudo-log-likelihood} subtraction and
the error from the Monte Carlo integration used to evaluate
$\hat\cali(\mu,\mu_{\rm mc})$ in the
resolution-efficiency correction procedure.  There is also a contribution to
the statistical error from the fluctuations in the background that {
are} not accounted for in the fit.  The first two can be reduced by
increasing the size of the Monte Carlo sample, but the latter error is
irreducible and must be included regardless of the size of the Monte
Carlo sample.

We have developed procedures for evaluating these three errors.  The
formulas used are collected in \Appx\ A.

\subsection{Background Level Estimation}
\label{sec:back}

The categorization of \bkgd s has been detailed in
Sec.~\ref{sec:bkgdcats}.  The 
 \comb\ and \pking\ backgrounds are estimated by fitting the measured
 $\Delta m=m_{K\pi^+\pi^-}-m_{K\pi}$ and \cosby\ distributions in
 data. The $\Delta m$ fit determines the combinatorial background.
 The \cosby\ distribution is then fit to a combination of signal,
 combinatorial background, ``$D^{**}$'' peaking and other peaking
 background using the binned likelihood fitting method of Barlow and
 Beeston~\cite{bandb}.  The shape of the 
\cosby-\dist s are taken from the \MCs.  In the fit, only the
signal fraction and the portion of the peaking background due to \dstst\ and
$D^*Xe\nu$ decays are free \param s.
The combinatorial fraction is set to the value obtained from
the $\Delta m$ fit and the other components of the peaking background
are scaled from the Monte Carlo. See  Sec.~\ref{sec:bkgdcatsdetail} for more 
details.

We apply the above procedure first to obtain the overall 
fractions for the combinatorial ($\fcomb$) and the \dstst\
($f_\dstst$) backgrounds.  These are obtained by fitting to the full
data sample.  The fraction of the consolidated remaining peaking
backgrounds ($\fother$), detailed in
Sec.~\ref{sec:bkgdcatsdetail}, is obtained by direct scaling from the
MC simulation.

The  weight needed to subtract the correct amount
of each background type is given by

\be
\label{eq:bkwt}
\Wtype=\frac{\ftype\times \Ndata}{\Ntype},
\ee
where type=$comb$, $\dstst$ or $other$ depending on the particular
\bkgd\ type being subtracted and $\Ntype$ is the number of Monte Carlo
events of that type available. 

The determination of $\fcomb$ and $f_{\dstst}$ is discussed in
more detail in the next two subsections.

\sssec{Determination of Level of Combinatorial \Bkgd}
\label{sec:combobkgd}

The combinatorial background is due to \dst\ candidates that were not
originally actual
\dst\ mesons. It includes candidates where the $D$ is properly reconstructed, but is paired with a random 
$\pi$; events in which the $\spi$  is in fact from the
decay $\dst\rightarrow D\spi$, but the $D$ is not properly
reconstructed; and candidates which are purely combinatoric (neither
the $D$ nor the $\spi$ are correctly reconstructed) -- {\it i.e.},
where any one or more of the three detected particles does not
originate from the decay chain $\Dst \rar  D \spi \rar (K \pi) \spi$.

To model this background we use the following functional form
\be
\label{eq:thrfunc}{
f(\Delta m)=N\left(1-\exp{\left(\frac{\Delta m-\Delta \mthresh }{s}\right)^{\kappa}}\right)\left(\frac{\Delta m}{\Delta \mthresh }\right)^{\gamma}},
\ee
where below the { $D^*$} threshold { $\Delta \mthresh $} we take
$f(\Delta m)$=0.  This is an extension of the commonly used threshold
function \cite{babarVcb}. To the single scale parameter $s$ the power
$\kappa$ and the factor $\lp{\Delta m} \over {\Delta \mthresh
}\rp^\gamma$ have been added as extra degrees of freedom.  This
extension allows us to obtain a better fit to the background and
allows the fit enough freedom to account for the uncertainty in the
background shape. Fixing $\kappa=1$ and $\gamma=0$ corresponds to the
usual unextended threshold function (further details of the necessity
of this extension are given in \Appx\ B).

The signal \dist\ is fit by three free Gaussian functions plus a
fixed tail of two wide Gaussians.  The tail is fixed by fitting pure
signal Monte Carlo as described in \Appx\ B.

The final fit to data is shown in Fig.~\ref{fig:deltamdata}.

\begin{figure}[htp]
\begin{center}
{\parbox{8cm}
  {\resizebox{8cm}{!}{\includegraphics{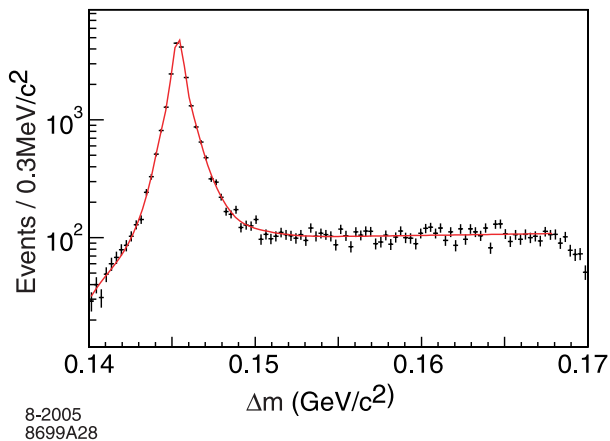}}
}}
\end{center}
\caption{
The fit (solid line) to the $\Delta m$ distribution for $\dst e\nu$
candidates as described in the text.  The distribution is shown
logarithmically to emphasize details of the functional fit, but
the actual level of background is more clearly seen in the linear data
plot shown in Fig.~\ref{fig:dmall1}.}
\label{fig:deltamdata}
\end{figure}

\sssec{Determination of Level of \Dstst\ \Bkgd}

To estimate the peaking background we use a mixture of Monte Carlo
predictions and a fit to the \cby\ distribution within the $D^*$
signal window $\Deltam = 0.143 -- 0.148\gevcc$.  We fit the \cby\
distribution for the signal and the background due to decays of
the type $B\rightarrow D^{*+}Xe\nu$ (called the ``$D^{**}$''
background, as defined in Sec.~\ref{sec:bkgdcatsdetail}). The
shapes of the \cby\ distributions for the signal and the
backgrounds are obtained from the \MCs. The backgrounds other than
the $D^{**}$ are fixed at values obtained by scaling from the
Monte Carlo by the appropriate luminosity ratios.  In the case of
fake electrons (which are mostly misidentified pions), we also
scale using the misidentification probabilities obtained from the
data control sample of pions (specifically from the \babar\
$\tau^\pm \rightarrow \pi^\pm \piplus \piminus \nu_{\tau}$
dataset).

Figure~\ref{fig:cby} shows the results of this fit to the full
data sample. The shading indicates the background source or
signal. Only two paramaters, the signal and the $D^{**}$
background fractions, are free in the fit. The combinatorial {
background} is input from the $\Delta m$ fit. The other peaking
backgrounds are scaled from the \MCs\ as described above.

\begin{figure}[htp]
\begin{center}
{\parbox{8cm}
  {\resizebox{8cm}{!}{
  \includegraphics{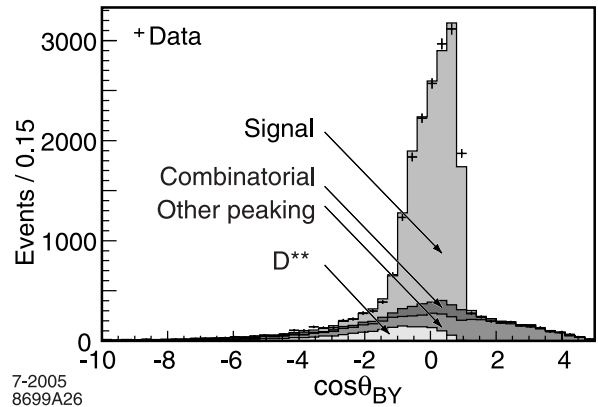}
  }
} 
}
\end{center}
\caption{
The \cby\ data distribution and the result of the fit with the
estimated signal and \bkgd\ contributions, as described in the text.
Shading indicates the various components in the fit.}
\label{fig:cby}
\end{figure}

For input to the form-factor fit we need the background fractions
within the \cby\ signal window ($|\cby|\leq 1.2$).  These are {
$\fcomb=5.33\pm 0.26\% $ for the combinatorial, $f_{D^{**}}=4.85\pm 0.35\% $ for the $D^{**}$ and
$\fother=7.03\pm 0.45\% $ for
peaking backgrounds other than the $D^{**}$}.

\sssec{{ Dependence of \Bkgd\ Levels on Kinematic Variables}}
 
In addition to obtaining overall background fractions, we perform our
$\Delta m$-\cosby\ fitting procedure in five bins for each of the
kinematic variables in the data. This allows extraction of the
dependence { of} combinatorial and \dstst\ backgrounds for each
kinematic variable.  We compare the ratio of fitted \bkgd\ yields in
data and MC, and parameterize the difference as a linear function of
the kinematic variables. This allows application of a correction term
to the weight for each Monte Carlo event of the form:
\be
\label{eq:correctwt}
\Wtype^{corr}=(1+\alpha_w(w-\left<{w}\right>_{\rm type}))
\ee
$$\times(1+\alpha_\ctl(\ctl-\left<{\ctl}\right>_{\rm type}))$$
$$\times(1+\alpha_\ctv(\ctv-\left<{\ctv}\right>_{\rm type}))$$
$$\times(1+\alpha_\chi(\chi-\left<{\chi}\right>_{\rm type})),$$
where the means for each type (which are calculated from the \MCs\ \dist s) are subtracted to keep the normalization independent of the slope.

To extract the dependence of combinatorial background on the kinematic
variables, the procedure of fitting four Gaussians for the signal
(discussed in \Appx\ B) and the extended threshold function for the
\comb\ \bkgd\ portion is repeated for five bins in each of the four
kinematic variables.

 To obtain the slopes needed in Eq.~(\ref{eq:correctwt}) we form the
 ratio of the background estimate in each bin to the number found in
 that bin in the \MCs\ and fit it to a straight line. All of the fits are
 displayed in in Fig.~\ref{fig:dmfitbins} \Appx\ B.  

This information is summarized and shown in Fig.~\ref{fig:comborat},
and the numerical results for the slopes are given in
Table~\ref{table:dmslopes} along with the means, which are needed to
do normalization independent weighting, as discussed above after
Eq.~(\ref{eq:correctwt}).

\begin{figure}[htp]
{\parbox{8cm}
  {\resizebox{8cm}{!}{
 \includegraphics{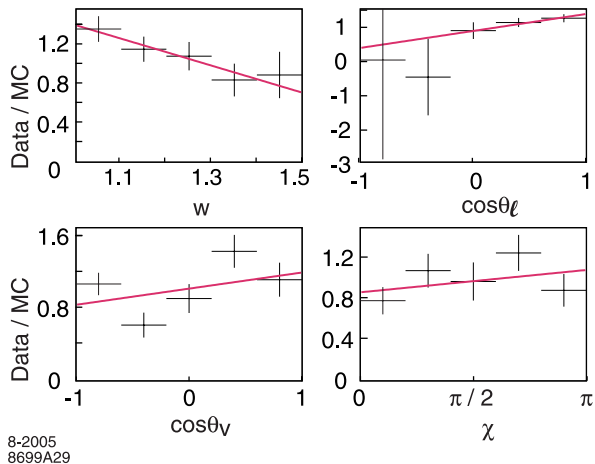}
}}}
\caption{\label{fig:comborat} 
Ratio of data to Monte Carlo events for combinatorial background for the four
kinematic variables $w$, \ctl, \ctv, and $\chi$, obtained from 5-bin fits to data and \MC\ in each \kv.  The lines are result of linear fits (see text).
}
\end{figure}

\begin{table}[htp]
\caption{\label{table:dmslopes}
The fitted slopes to the distributions shown in
Fig.~\ref{fig:comborat} of the ratio of data to Monte Carlo for the
combinatorial background in the four kinematic variables $w$, \ctl,
\ctv and $\chi$, and the mean values of the Monte Carlo
distributions for these variables.}
\begin{center}
\begin{tabular}{ c c c} 
\hline\hline
 Variable\hfill & \ Slope & Mean \\ 
\hline
 $w$ & $ \hphantom{+}0.51\pm 0.50$ & 1.23 \\
\ctl      & $  \hphantom{+}0.05\pm 0.11$   &     0.31     \\
\ctv    & $-0.20\pm 0.09$  &     0.00     \\
$\chi$         & $-0.13\pm 0.06$   &     1.48     \\
\hline\hline
\end{tabular}
\end{center}
\end{table}

As in the combinatorial case, we repeat the extraction of the \dstst\
component of the peaking backgrounds in five bins in each of the four
kinematic variables. The results of all these \cby\ fits are shown in
{ Fig.}~\ref{fig:cbybins} in \Appx\ B. This is summarized in the
linear fit for data to Monte Carlo events ratio for the $D^{**}$ background
(the component free in the fit), which is shown in {
Fig.}~\ref{fig:cbyslopes}, with the numerical results in
Table~\ref{table:cbyslopes}.

The linear fits in Fig.~\ref{fig:cbyslopes} give the slopes used in
weighting the backgrounds using Eq.~(\ref{eq:correctwt}).

We use the
central slope values to reweight our background subtraction and
propagate the errors into the systematic uncertainty by varying the
slopes within their errors.

\begin{figure}[htp]
\begin{center}
{\parbox{8cm}
  {\resizebox{8cm}{!}{
  \includegraphics{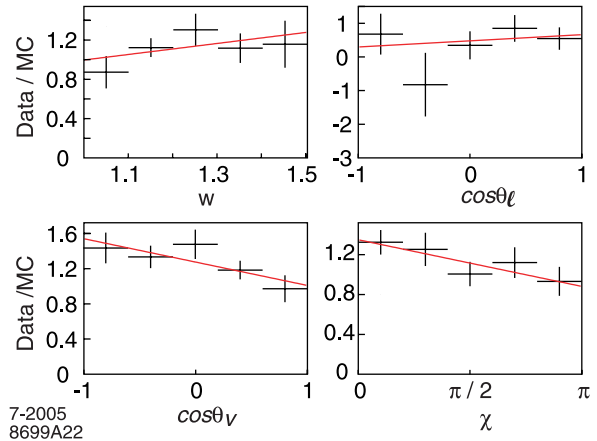}
} 
} }
\end{center}
\caption{\label{fig:cbyslopes}
Ratio of data to Monte Carlo events for \Dstst\ background for the four
kinematic variables $w$, \ctl, \ctv, and $\chi$, obtained from 5-bin fits to data and \MC\ in each \kv.   The lines are result of linear fits (see text) }
\end{figure}

\begin{table}[htp]
\caption{\label{table:cbyslopes}
The fitted slopes to the \Dstst\ distributions shown in
Fig.~\ref{fig:cbyslopes} of the ratio of data to Monte Carlo for the
combinatorial background in the four kinematic variables $w$, \ctl,
\ctv and $\chi$, and the mean values of the Monte Carlo
distributions for these variables.}
\begin{center}
\begin{tabular}{ c c c } 
\hline\hline
 Variable\hfill & \ Slope & Mean \\ 
\hline
 $w$ & $ -1.35\pm 0.52$ & $\hphantom{+} 1.28$ \\
\ctl      & $\hphantom{+}0.14\pm 0.29$   &   $\hphantom{+} 0.35$     \\
\ctv    & $\hphantom{+}0.18\pm 0.12$  &     $-0.11$    \\
$\chi$         & \hphantom{+}$0.07\pm 0.08$   &    $\hphantom{+} 1.53$     \\
\hline\hline
\end{tabular}
\end{center}
\end{table}

Further details of the fits and estimates for the background fractions and
weights are given in \Appx\ B.

\ssec{$\chi$ Averaging}

Few data events arising from true signal will be found at values of
the kinematic variables for which the PDF goes to zero.  However,
there is no such limitation on the number of data events arising from
backgrounds that may fall at these points.  Because of this, when the
logarithm of the PDF is evaluated for these points as part of the
background-subtracted computation of the log-\likeld\, it develops
spuriously large (negative) values, driving the fit far astray from the
actual minima.

To deal with these spurious zeros we deliberately introduce a small
amount of artificial resolution that prevents the PDF from going to
zero. This resolution is chosen to be small compared to the natural
resolution of our kinematic variable reconstruction
(see Fig.~\ref{fig:kinvarres}) and so has little effect on the sensitivity
of our measurement. Any small bias introduced is removed by the { resolution-efficiency
correction}
method in the same manner as that due to naturally occurring
resolution.

We implement this artificial resolution by evaluating the PDF for
all data points (including the data, Monte Carlo-selected
\bkgd, and integration samples, for consistency) at values of the phase
space not exactly {\it at} the given point, but offset from it on both
sides in $\chi$ by $\pm 0.1$. The \kv\ $\chi$ is chosen because it is
at particular discrete, parameter-dependent values of $\chi$ that the theoretical PDF is
zero.  We then take the values of the PDF for these two
offset points and average them. This average is used as the value for
the given point in evaluating its contribution to the likelihood.

Testing on \MCs\ shows $\chi$-averaging does not bias the fit
values, nor increase the errors.  Varying the value by which we move
$\chi$ we find that $\pm 0.1$ is small
enough to not bias the fit values, and large enough to eliminate the
spurious zero problems fully, so we evaluate all fit results with this
offset.

\section{Results}
\label{sec:Physics}

\ssec{Free \Param\ and Baseline Fits}

We perform the baseline fit (see Sec.~\ref{sec:hqsrships}) taking $R_1$ and $R_2$ to be independent of $w$ and
use the single parameter form factor description of Ref.~\cite{CLN} given in Eq.~(\ref{eq:clnff}).

We find

\bea
\label{eq:r1r2rhosq} 
\rone=1.396\pm 0.060\pm 0.035, \nonumber \\
\rtwo=0.885\pm 0.040\pm 0.022,  \\
\rhosq=1.145\pm 0.059 \pm 0.030, \nonumber
\eea
where the first {uncertainties given are due to the limitations of the data statistics and the second to limitations of statistics of the Monte Carlo simulations}.  Systematic uncertainties are discussed
in Sec.~\ref{sec:Systematics} below.  The errors are highly
correlated.  The error matrix for the full statistical error for
\rone, \rtwo, and \rhosq\ (including Monte Carlo \statl) is:

\bcenter
\begin{verbatim}  
       0.00479  -0.00243   0.00277 
      -0.00243   0.00207  -0.00231
       0.00277  -0.00231   0.00447
       \end{verbatim}
\ecenter

The correlations are

\bea
\eta_{R_1-R_2}= -0.77,  \nonumber \\
\eta_{R_1-\rho^2}= +0.60, \\
\eta_{R_2-\rho^2}= -0.76  \nonumber.
\eea

As we do not yet have enough \statl\ sensitivity to fit for the $w$-dependence of
$R_1(w)$ and $R_2(w)$, we consider instead the effect of the
theoretically predicted dependence on the result.  Parameterizing this
dependence as follows

\be
\label{r1w}
R_1(w)=R_1+\alpha_1(w-1)+\beta_1(w-1)^2,
\ee
\be
\label{r2w}
R_2(w)=R_2+\alpha_2(w-1)+\beta_2(w-1)^2,
\ee
and inserting these $w$-dependent forms into the PDF (with fixed $\alpha_i, \beta_i$ from 
the theoretical predictions) and fitting for the constant terms \rone\ and \rtwo\, we find the
results given in Table \ref{table:theorydep}.


\begin{table*}[ht]
\begin{center}
\begin{tabular}{ cl l l l l l l} 
\hline\hline
 Reference\hfill & \ $\alpha_1 $ & $\beta_1$ & $\alpha_2$ & $\beta_2$
 & $R_1$ & $R_2$ & $\rho^2$ \\ 
\hline
Baseline                               & $\hphantom{+}0.0$ &     0.0      & 0.0   & $\hphantom{+}0.0$ & 1.40 & 0.89 & 1.15 \\

Caprini-Lellouch-Neubert\cite{CLN}     & $-0.12$           &     0.05     & 0.11         & $-0.06$ & 1.42 & 0.87  & 1.12 \\

Ligeti-Grinstein\cite{LigetiGrinstein} & $-0.10$           &     0.0      & 0.09         & $\hphantom{+}0.0$ & 1.42 & 0.87 & 1.11 \\

 Neubert\cite{NeubertPhysReport}       & $-0.22$ & 0.09 & 0.15 & $-0.04$                      & $1.45$ & 0.86 & 1.09 \\
\hline\hline
\end{tabular}
\end{center}
{\caption{\label{table:theorydep} Dependence of form-factor parameters
on theoretical assumptions about slope ($\alpha$) and curvature
($\beta$) of $R_1$ and $R_2$ $w$-dependence. See Eqs. (\ref{r1w}) and
(\ref{r2w}).  }}
\end{table*}

These theoretical variations in $R_1(w)$ and $R_2(w)$ yield slightly
larger values for $R_1$ and slightly smaller values for
$R_2$. However, this is presumably only because they all decrease the
mean of $R_1(w)$ and increase the mean of $R_2(w)$ when averaged over
the $w$ spectrum.

\ssec{Higher-Order Free \Param\ Fits}
\label{sec:hofits}

We performed a study using a large MC sample in order to ascertain
whether it was also useful for us to do free fits to \haone\ with up to five
\param s (see Eq.~(\ref{eq:ffexpd})) for a sample generated with a particular form (the CLN form).
To this end, we created a MC sample of a million events generated with
the ansatz of Eq.~(\ref{eq:clnff}) with no acceptance cuts or
resolution smearing applied.  We then attempted to fit this using four
parameters (\rone,\rtwo, and \coef s to \order$(w-1)^2$ in \haone) and five
parameters (\rone,\rtwo, and \coef s to \order$(w-1)^3$ in \haone).  The fit
results for the slope and curvature (for the four \param\ fit) and
slope, curvature, and cubic term coefficient (for the five \param\
fit) were far from the generated values corresponding to the $(w-1)^n$
expansion of the input (Eq.~\ref{eq:clnff}).  

This problem arises because the fit is attempting to compensate for
the missing higher-order terms. As a result the coefficients of
$(w-1)$ or $(w-1)^2$ returned by the fit cannot be cleanly interpreted
in terms of the theoretical expectations or interpreted in terms of
limits such as those given in
\cite{oliver,oliver2}.

Thus, while either the CLN or free polynomial expansion are {\it a priori}
valid, the former is preferred from theoretical constraints, and we
see from the above that it is not useful to attempt to fit a sample
with a free expansion to a reasonably low expansion order and then
interpret the result in terms of CLN coefficients.

\ssec{Iterative Convergence Studies}
\label{sec:iterativeconv}

We have studied the convergence of the likelihood maximization
procedure described in Sec. VI.A.  Varying the initial seed values by
as much as 50\% of the final values still leads to convergence to the
same final values within 0.0005 for each parameter within five
iterations.



\section{Goodness-of-Fit}
\label{sec:gof}
To assess whether the results of the fit reproduce the
distribution of the kinematic variables in the data, we use a binned
$\chi^2$ method of estimating \gof.  Chernoff and
Lehmann~\cite{chernoff} have shown that while a binned $\chi^2$ will
in general have a wider than expected distribution when the free
parameters are optimized using a likelihood fit, the effect is small
for a large number of bins.  Therefore we adopt the binned $\chi^2$ as
our primary goodness-of-fit test.  

Since we do not have an explicit form for the acceptance-corrected PDF
($\calF$) of the four reconstructed variables $w$, $\ctl$, $\ctv$
and $\chi$, we reweight the MC sample to construct the distributions
expected from our measured parameters. That is, the contribution of
the signal to a bin is given by the MC event sum

\be
\label{eq:wtsum}
n_{\rm signal}=\sum_{i \in {\rm MC}} {W^{(i)}_{\rm signal}},
\ee
where 
\be
\label{eq:wt}
W^{(i)}_{\rm signal}=f_{signal}\times N_{\rm data}\times\frac{W_i}{\sum W_i}
\ee
and in this case $W_i=\frac{F(x_i;\mu)}{F(x_i;\mu_{mc})}$ is the weight
needed to modify the distributions from those generated with
$\mu_{mc}$ ($R_1=1.18$, $R_2=0.72$, and $\rho^2=0.92$), to those
obtained from this analysis and $\fsignal=1-f_\dstst-\fother-\fcomb$ 
is the signal fraction.

For the background we use the same weighting procedure (see
Eqs.~(\ref{eq:bkwt}) and (\ref{eq:correctwt})) used in the fit. Using
these weighting procedures the normalizations of the data and
reweighted distributions match by construction.

We consider two types of goodness-of-fit: 
\bi
\item A four-\diml\ binned $\chi^2$ based on a total of
1296 bins, that is, $6\times 6\times 6\times 6$ bins, six each for the four kinematic variables 
\item One-dimensional
projected distributions of $w$, $\ctl$, $\ctv$, and $\chi$, as well as
the distributions of the CM electron momentum ($p^*_\ell$) and the transverse
momentum ($p_t$) of the slow pion from the $D^*$ decay.
\ei

\ssec{Four-\diml\ Binned \boldmath{$\chi^2$}}
Six bins were chosen for the $\chi^2$ to make sure we have adequate
statistics in each bin while still having sensitivity to the shape
predicted by Eq.~(\ref{eq:thlpdf}) combined with detector and event
selection acceptance.  The kinematically empty regions shown for 
Monte Carlo events in Fig.~\ref{fig:empty}a for $w\geq 1.08$ and
backward $\ctl$ (due to the electron momentum cut) and in
Fig.~\ref{fig:empty}b for small $w$ and forward $\ctv$ (due to slow
pion acceptance) are excluded by simple cuts on minimum $\ctl$ and
maximum $\ctv$ as functions of $w$, respectively.


\begin{figure}[hb]
\begin{center}
{\parbox{7cm}
 {\resizebox{7cm}{!}{\includegraphics{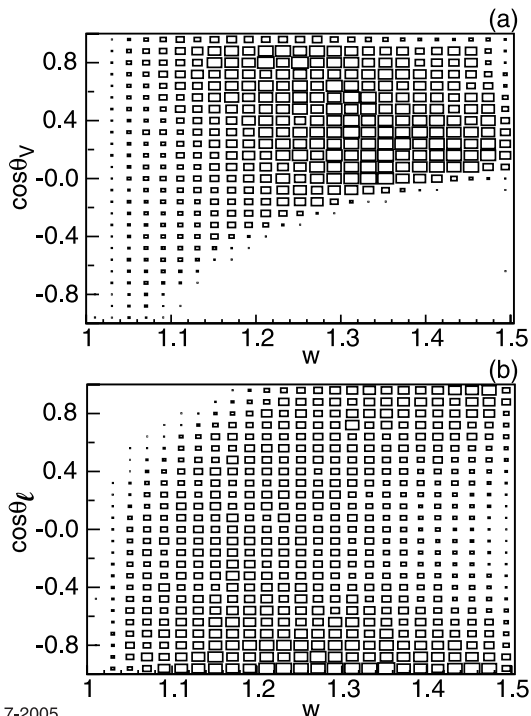}} 
}}
\end{center}
\caption{\label{fig:empty} 
Plots showing empty regions (due primarily to cuts) in the phase space of the kinematic
variables $w$, \ctl, and \ctv\ using the Monte Carlo sample: (a)
shows \ctl\ vs. $w$, where the high $w$, low \ctl\ region is removed in the \gof\ calculation; (b) shows \ctv\ vs. $w$, the high \ctv, low $w$ region is removed in the \gof\ calculation.  
}
\end{figure}

The $\chi^2$ is given by
\be
\chi^2=\sum_i\left(\frac{n_i-m_i}{\sigmadif} \right)^2,
\label{eq:bindchisq}
\ee
where the sum is over the 1296 bins, $n_i$ is the number of data
events in bin $i$, $m_i$ is the weighted number of Monte Carlo events
and $\sigmadif$ is the error on the difference $n_i - m_i$. 
There is, of course, an error on
our estimate of the error $\sigmadif$ too, that makes the $\chi^2$ 
contributions from low
statictics bins vary widely. Thus, estimating $\sigmadif$ is
critical to obtaining a reasonably behaved $\chi^2$
distribution. 

Using a simple \MCs\ divided into 1296 bins that
roughly mimics the number of low statistics bins in our dataset, 
we can achieve a
flat $\chi^2$ probability distribution if we base the error on the
best estimate of the number of events in each bin computed using the
overall data-to-Monte-Carlo events ratio $\rmc$. 
That is $m_i$ is given by
muliplying the number of MC events $n^{\rm mc}_i$ by $\rmc$. 

Specifically, we use $n^{\rm best}_i=f n_i+(1-f)m_i$ where $f \equiv
{\rmc} / {(1+\rmc)}$. The difference error $\sigmadif$ is obtained by
using $\sqrt{n^{\rm best}_i}$ for the contribution from the data and
by scaling the error on $m_i$ by $\sqrt{n^{\rm best}_i/m_i}$.  Thus,
we use the hypothesis we are testing (that the weighted Monte Carlo
follows the same distribution as the data) to obtain the best possible
estimate of $\sigmadif$ by using the best estimators we can obtain for
the number of data events ($n^{\rm best}_i$) and for the number of MC
events ($n^{\rm best}_i/\rmc$) to estimate their respective
contributions to the error. This procedure reduces the number of bins
where the error estimate fluctuates low leading to artificially high
$\chi^2$ contributions.

For the baseline fit we find
\be
\label{eq:basechisq}
\chi^2 =1336.66
\ee
for 1292 degrees-of-freedom (1296 bins minus three free parameters and
one for normalization), which yields a $\chi^2$ probability of $19\%$
indicating the fitted parameters yield a good description of the full
four-dimensional distribution of the data (note that we keep two
decimal significant digits in $\chi^2$ since every one unit change in
correponds to a full one sigma shift in the fit parameter).

\ssec{One-\diml\ Binned  Projection Plots}

For a more detailed examination of how well we fit the data, we turn
to one-\diml\ projection plots.  Fig.~\ref{fig:uncutkinvar} shows the
weighted Monte Carlo (histogram) overlaid on the {
background-subtracted} data. The \bkgd s are subtracted in the
one-\diml\ analog to the method used in the fit, as detailed in
Sec.~\ref{sec:back}.

The difference divided by its error (``pull'') is shown below each
plot. The agreement is very good. The $\chi^2$ and the corresponding
probability for each variable is summarized in
Table~\ref{tbl:chisqs}. We take the number of degrees-of-freedom
(\ndof) to be the number of bins minus one for purposes of
estimating the probabilities.

To test our sensitivity to the interference terms in
Eq.~(\ref{eq:thlpdf}) we make the same kind of comparisons in six bins
of the angle $\chi$ for the variables \ctl and \ctv.
Since the normalization is not fixed in these plots the number of degrees-of-freedom is 20. 
The \ctl plots are shown in
Fig.~\ref{fig:cutctl}. The \ctv\ plots are in Fig.~\ref{fig:cutctv}.
In Table~\ref{tbl:chisqctlctv} we collect the $\chi^2$ and its probability
for each plot. Again the agreement is excellent indicating that the
fit succeeds well in reproducing the details of the distribution.

In these plots we can also see the effect of isolating specific
portions of the PDF -- {\it e.g.}, in the very low $\chi$ region, $\cos
\chi \rar 1$ and we obtain the modulation effect of the $\stv \ctv$ in
the fifth term of the PDF, seen in the top left plot in
Fig.~\ref{fig:cutctv}.  On the other hand, in the high $\chi$ region,
$\cos \chi \rar -1$ and this effect flips sign due to the sixth term,
as can be clearly seen in the bottom right plot.  (These types of
effects also occur in Fig.~\ref{fig:cutctl} but they are masked by the
effect of the \pstarl\ cut, which suppresses the low \ctl\ region
strongly in that case, so the modulations are more difficult to
observe).  Much of the information content of our data is encoded in
the interference terms, so it is important to ensure that we
reproduce their effect accurately.

\begin{table}[ht]
\caption{\label{tbl:chisqs} $\chisq$ and $\chisq$-probability 
for kinematic variable projections and electron momentum. The number of
bins in these histograms is either 20 or 16.}. 
\begin{center}
\begin{tabular}{ccc}
\hline\hline
variable & $\chi^2$/\ndof  & $\chi^2$ probability   \\
\hline
$w$ & 22.0/19 & 28\%  \\
$\ctl$ & 23.0/19 & 24\% \\
$\ctv$ & 31.8/19 & 3.3\%  \\
$\chi $ & 13.0/19 & 84\%  \\
$p^*_\ell$ & 17.3/19  & 57\%  \\
$p_t$ & 9.0/15 & 88\%  \\
\hline\hline
\end{tabular}
\end{center}
\end{table}

\begin{table}[ht]
{\caption{\label{tbl:chisqctlctv} $\chisq$ and $\chisq$-probability
for \ctl\ and \ctv\ overlay plots in six bins of the \kv\ $\chi$.  These
values correspond to the numerical evaluation of one-\diml\ \chisq\ of
Figs.~\ref{fig:cutctl}~and~\ref{fig:cutctv}.  }}
\begin{center}
\begin{tabular}{cccc}
\hline\hline
variable & $\chi$ cut & $\chi^2$/\ndof  & $\chi^2$ probability   \\
\hline
\ctl & $0\leq\chi\le\frac{\pi}{6}$ & 18.4/20 & 56\%  \\
$\ctl$ & $\frac{\pi}{6}\leq\chi\le\frac{2\pi}{6}$ & 19.3/20 & 50\% \\
$\ctl$ &  $\frac{2\pi}{6}\leq\chi\le\frac{3\pi}{6}$ & 29.6/20 & 7.7\%  \\
$\ctl$ &  $\frac{3\pi}{6}\leq\chi\le\frac{4\pi}{6}$ & 17.9/20 & 59\%  \\
$\ctl$ &  $\frac{4\pi}{6}\leq\chi\le\frac{5\pi}{6}$ & 23.9/20  & 25\%  \\
$\ctl$ &  $\frac{5\pi}{6}\leq\chi\le\pi$ & 12.4/20 & 90\%  \\
\hline
$\ctv$ &  $0\leq\chi\le\frac{\pi}{6}$ & 19.5/20 & 49\%  \\
$\ctv$ &  $\frac{\pi}{6}\leq\chi\le\frac{2\pi}{6}$ & 26.7/20 & 14\% \\
$\ctv$ &  $\frac{2\pi}{6}\leq\chi\le\frac{3\pi}{6}$ & 10.8/20 & 95\%  \\
$\ctv$ &  $\frac{3\pi}{6}\leq\chi\le\frac{4\pi}{6}$ & 20.1/20 & 45\%  \\
$\ctv$ & $\frac{4\pi}{6}\leq\chi\le\frac{5\pi}{6}$ & 27.4/20  & 12\%  \\
$\ctv$ &  $\frac{5\pi}{6}\leq\chi\le\pi$ & 25.3/20 & 19\%  \\
\hline
\hline\hline
\end{tabular}
\end{center}
\end{table}


\begin{figure}[ht]
\begin{center}
{\parbox{8cm}
  {\resizebox{8cm}{!}{\includegraphics{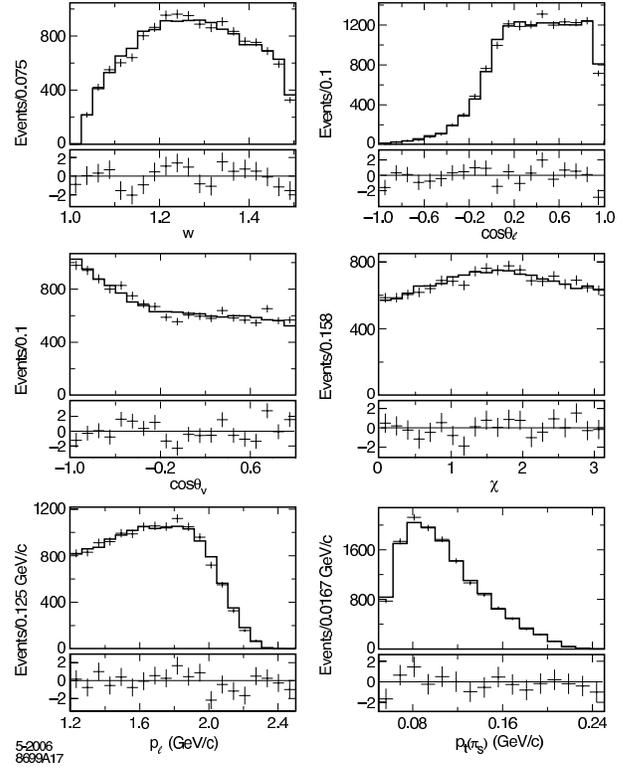}} 
}}
\end{center}
\caption{\label{fig:uncutkinvar} { Background-subtracted} data (points) 
overlaid on Monte Carlo (histograms) for all four \kv\ distributions
and for the electron momentum (\pstarl) and transverse momentum
($p_t$) of the slow pion for our best fit.  In the bottom panel of
each figure is shown the pull (difference over error) plot.  The red
line in the pull plot is not a fit, but is the line at zero shown for
comparison (similarly for all following pull plots).  }
\end{figure}

\begin{figure}[hb]
\begin{center}
{\parbox{8cm}
  {\resizebox{8cm}{!}{\includegraphics{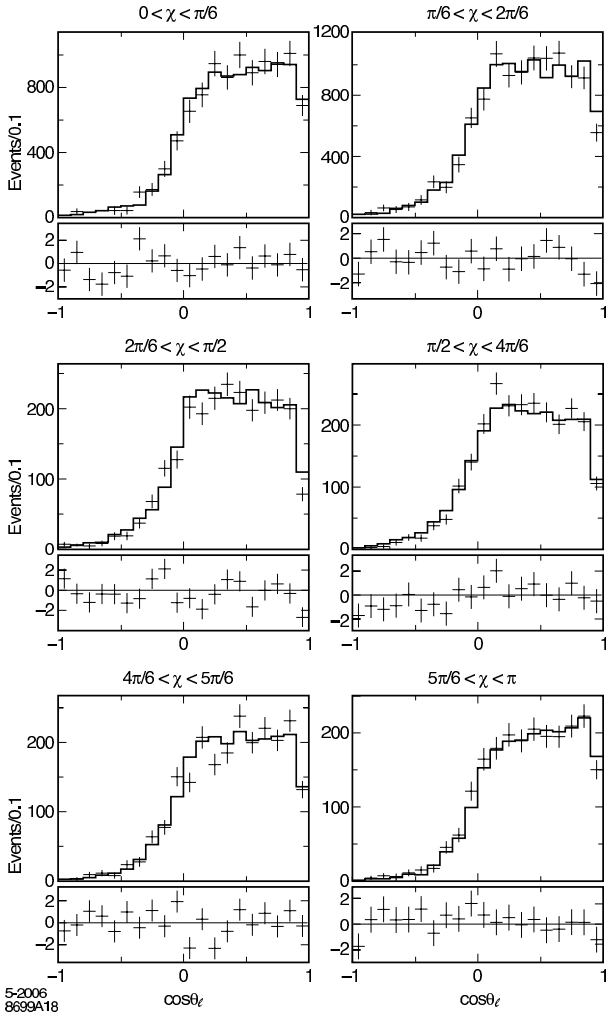}} 
}}
\end{center}
\caption{\label{fig:cutctl} Background-subtracted data (points) 
overlaid on Monte Carlo (histograms) and pull plots
for \ctl for six $\chi$ cuts.
\newline
}
\end{figure}


\begin{figure}[hb]
\begin{center}
{\parbox{8cm}
  {\resizebox{8cm}{!}{\includegraphics{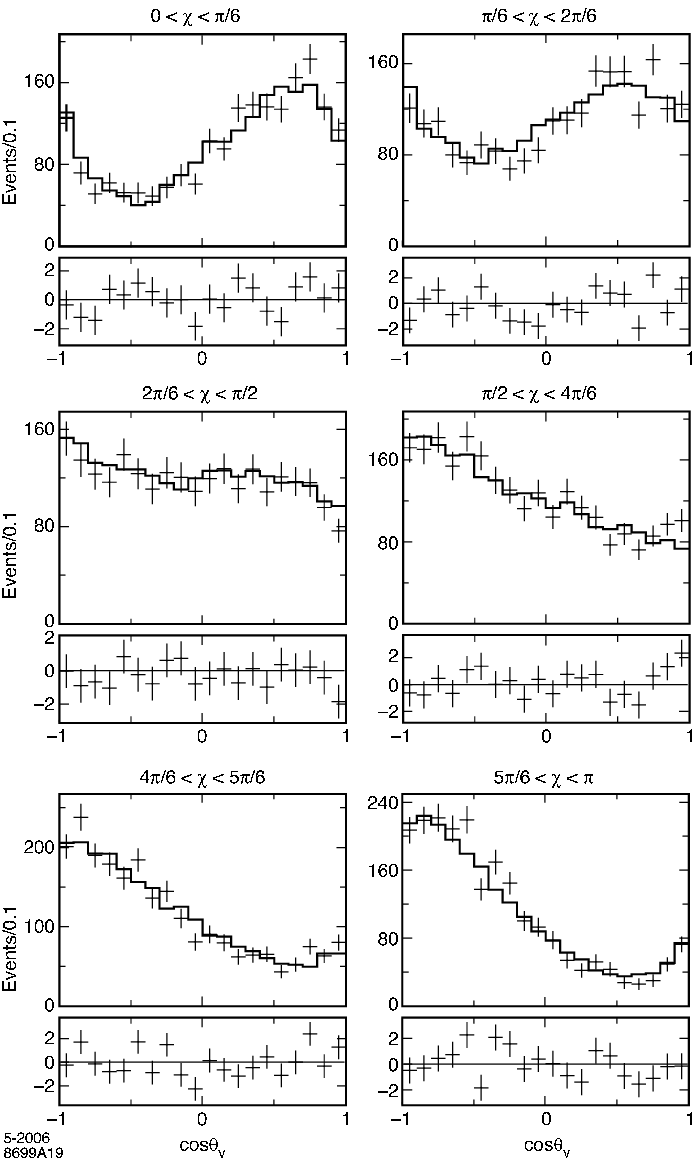}} 
}}
\end{center}
\caption{\label{fig:cutctv} Background-subtracted data (points) 
overlaid on Monte Carlo (histograms) and pull plots for \ctv\ for six
$\chi$ bins.  These plots illustrate our sensitivity to interference
effects.  }
\end{figure}

\section{Systematic Studies}
\label{sec:Systematics}
\label{sec:systerrs}

The systematic uncertainties on the three parameters $R_1$, $R_2$ and
$\rho^2$ of the baseline fit (see Sec.~\ref{sec:hqsrships}) are
summarized in Table~\ref{table:CLNsysterrs}. The systematic errors are
comparable for fits which include variation in $w$ of \rone\ and \rtwo, such as those shown in Sec.~\ref{sec:Summary}.

A major source of  systematic error arises from the MC simulation,
specifically, from the modeling of the detector resolution and
efficiency. We are particularly concerned with the efficiency for
reconstructing the low-momentum charged pion from \Dstarplus\ decays.

  Further, how well the background event generation is modeled, {\it
  e.g.} how close the branching fractions in the event generator are
  to measured ones, affects the distribution of background we
  subtract.  The kinematic-variable dependence of the backgrounds,
  however, is not taken from the Monte Carlo but is measured in the
  data under the assumption that the difference between the Monte
  Carlo and the data can be represented by a linear kinematic-variable
  dependence.  The uncertainty in this measurement also contributes to
  the systematic errors. We are also able to check our background
  estimates using the goodness-of-fit as shown in
  Section~\ref{sec:gof}.

Our systematic errors fall broadly into two categories:

 \begin{itemize}
 \item {A: Detector simulation, {\it i.e.}, the accuracy with which the 
Monte Carlo reproduces the resolution and efficiency of the detector. 
We are most sensitive to how well it models the complex 
process of detecting and measuring {\rncversion low-momentum} pions.}
 \item  {B: Simulation of $B$ decays and \bkgd, {\it i.e.}, the accuracy with which our event 
generation models the signal and background distributions and
how well we correct for differences between data and Monte Carlo.
}
 \end{itemize}

\subsection{Detector Simulation}
Extensive studies of the simulation of the detector response,
including careful examination of track reconstruction and
particle identification efficiencies, have been performed using selected data
control samples.  Adjustments for known simulation deficiencies are
used in investigating and evaluating the systematic errors.  Form-factor 
measurements are insensitive to overall normalization errors.
Thus differences in the efficiencies that are independent of the fit
variables do  not affect the results, but MC/data differences that vary
as a function of these variables are of concern.

To assess the uncertainties due to differences in shape rather than
normalization (Sec.~\ref{sec:trking}--Sec.\ref{sec:fsr}), we vary
the efficiency corrections, reweight the Monte Carlo samples with
these modified corrections, and rerun the fit to the \kv\ data.  Finally,
we take the difference between the results of these fits and
those obtained with the nominal Monte Carlo simulation as an estimate
of the systematic error on the parameters.  This procedure is repeated
for each source of detector-based systematic uncertainty. The
individual uncertainties are added in quadrature to obtain the total
systematic error.

\subsubsection{Charged Particle Identification (PID)}
\label{sec:piderr}

Using data and Monte Carlo simulated control samples we have studied
the difference in particle identification efficiency
$\varepsilon_{pid}$ between data and Monte Carlo. We calculate a
correction factor, $ \varepsilon^{(data)}_{pid} /
\varepsilon^{(mc)}_{pid} $ , as a function of momentum.  For electrons
the correction factors vary from 0.991 to 1.008 over the momentum
range from $1.2 \gevc$ to $2.5
\gevc$.  We assess the impact of the uncertainty in these corrections
by approximating their momentum dependence by linear functions and varying
the size of the small slope of these functions by its one-sigma
uncertainty.  The observed deviations from the default fit are $\Delta
R_1=0.0064$, $\Delta R_2=0.0052$ and $\Delta \rhosq=-0.0016$ for the
increased slope and $-0.0032$, $-0.0031$, $+0.0009$ for the decreased slope.
We take half of the difference as the systematic error from this
source.  Since the momentum dependence is not a monotonic function,
this procedure slightly overestimates the uncertainty.

For kaon identification we employ the same procedure. The observed
variations are significantly smaller.

The probability of misidentifying the charged hadrons $\pi^{\pm}$,
$K^{\pm}$, $p^{\pm}$, as electrons is small, less than $0.2\%$ in
the momentum range $1.2-2.5 \gevc$.  Since a variation of the peaking
background by 9\% results in a very modest change in the fit results,
and since the fraction of this background originating from hadrons
misidentified as electrons is small, we conclude that the
uncertainty in the hadron misidentification rate is negligible.

The misidentification rate of pions as kaons as a function of pion
momentum ranges from a few tenths of a percent to almost
$5\%$. However, pion misidentification is well simulated by the Monte
Carlo and thus should have little impact on the fit
results. Furthermore, the main consequence of pion misidentification
is to increase the combinatorial background. Since we estimate the
combinatorial background from a fit to the measured $\Delta m$
distribution, we are not dependent on the Monte Carlo to assess the
size of this background. We conclude that the uncertainty in the pion
misidentification rate has little impact on the fit results.

\subsubsection{Charged Particle Tracking}
\label{sec:trking} 

The difference between data and \MC\ for the tracking efficiency for
electrons, charged kaons and pions decreases roughly linearly as a
function of momentum.  Analogously to the method described for the PID
error of Sec.~\ref{sec:piderr}, we vary this linear dependence on the
particle momentum in the Monte Carlo simulation and take the small
deviation as the error from this source.  It is a small contribution
to the total \systerr.

\subsubsection{ Slow Pion Reconstruction }
\label{sec:spierr}

The efficiency for reconstructing the low-momentum charged pion
($\pi_s$) from the decay $D^{*+} \rar \Dzero \spi$ is a major source of systematic
error. Because of the small energy release in \Dstar\ decays, this
pion is emitted in approximately the same direction as the parent
\Dstar\ and its momentum in reconstructed events varies from approximately 50 to 400
\mevc\ in the laboratory frame, peaking at $\sim 100 \mevc$, and with
a long tail extending out to 400 \mevc.  Since $w=E_{D^*}/m_{D^*}$ in
the $B$ rest frame, the $\pi_s$ momentum is correlated with $w$ and thus
its momentum-dependent efficiency impacts especially the measurement
of $\rho^2$.

The uncertainty due to the low-momentum tracking efficiency is
evaluated differently from other tracking errors because such
low-momentum tracks do not traverse the whole drift chamber.  Their
detection and measurement depends primarily on the silicon vertex
tracker.  To study this efficiency as a function of $p_{\pi_s}$ we use
a large sample of $\dsp\ra\Dz \pi_s^+$ decays selected from all
events and measure the distributions of the helicity angle of
$\pi_s^+$ ($\theta_{\pi_s})$ in the $D^{*+}$ rest frame as a function of the
$D^{*+}$ momentum.

We parameterize the $\pi_s$ efficiency as function of its
momentum using the form
\be
\label{eq:epis}
\varepsilon (p_{\pi_s})=\varepsilon_{\rm max}\left(1-\frac{1}{1+\beta(p_{\pi_s}-p_0)}\right)
\ee
with $p_0$ being the threshold momentum and $\beta$ controlling the
rapidity with which the efficiency rises above threshold.  For
$p_{\pi_s}<p_0$ we set $\varepsilon$ to zero.  We fit the data and
Monte Carlo helicity angle distributions for $\beta$, $p_0$ and the
coefficient of the $\cos^2\theta_{\pi_s}$ term (allowed because the \Dstar\ is in general polarized to varying degrees in each momentum bin, see
Sec.~\ref{sec:slowpi}) to obtain efficiency functions for the data and
the Monte Carlo.  The helicity method only determines the relative
momentum dependence of the efficiency. The absolute dependency is not
relevant for this analysis.

To assess the systematic uncertainty due to the $\pi_s$ efficiency
(\spieff) we weight the Monte Carlo simulation by the ratio of the
data function to the Monte Carlo function and assign the observed
shifts in the fitted values for $R_1$, $R_2$ and $\rhosq$ as
systematic errors.  As expected, $\rhosq$ is the most sensitive
($\Delta\rhosq=0.018$) to
\spieff\ since, of all the \kv s, \spieff\ most strongly affects the
shape of the $w$-distribution, which is the \kv\ that has the
strongest impact on the extracted \rhosq\ value.

\subsection{Event Simulation}

\subsubsection{Final State Radiation \rncdelete }
\label{sec:fsr}

Final state radiation, primarily from electrons in the decay chain,
lowers the momenta and to a lesser degree changes the angles of
detected particles.  Though a physics effect, {\rncversion final-state
radiation} acts much like a resolution -- it smears the kinematic
variables.  We simulate the emission spectrum of radiative photons
using PHOTOS~\cite{Was}, so the resolution-efficiency correction
procedure properly corrects for {\rncversion final-state radiation} to
the extent that PHOTOS models it correctly.

To test the sensitivity to uncertainty in the simulation of {\rncversion final-state radiation} we
evaluate the shifts in the fitted values of $R_1$, $R_2$ and $\rhosq$
between fits done with and without {\rncversion final-state radiation} corrections.  We assume an
uncertainty of 30\% in the simulated photon emission and thus take
$\sim 1/3$ of the observed shifts ({\rncversion final-state radiation} on vs. {\rncversion final-state radiation} off) of
$0.0129$, $0.0067$ and $0.0039$ as an estimate of the systematic
uncertainty.

\sssec{Peaking Background Mixture Uncertainty  }
\label{sec:backall}

The modeling of the peaking background depends on the knowledge of the
branching fractions for the mixture of semileptonic $B$ decay modes
that make up this background.  These branching ratios and the \FF s
for these modes are not very well measured.  To estimate the
uncertainty associated with these branching fractions, we use a
one-sigma variation of the procedure discussed in
Section~\ref{sec:back}, {\it viz.}: we first fit the
\cosby\ \dist s with a sum of functions where
the signal fraction  and the contribution due to
the \dstst\ \bkgd\ are allowed to float.  This fit is done in five
{\rncversion equal}-sized bins for each \kv, over the full kinematically allowed
range of the variable, in both data and Monte Carlo.  We then take
data-to-Monte Carlo ratios of the fit results per bin  (shown in 
Fig.(\ref{fig:cbyslopes})), which we then 
fit to a linear function of the kinematic variables.  While for
the central value we use this function to weight the
\Dstst\ \bkgd\ itself before it is subtracted from the data, for the
error we vary the weighting function up and down by its {\rncversion one-sigma}
slope uncertainty as shown in Table~\ref{table:cbyslopes}.  The \diff\
of the original fit minus the result of this fit gives us an estimate
of the error due to the uncertainty in the shape of the \dstst\
background for each \kv.  We add the uncertainties for each \kv\ in
quadrature to give us the total error due to this source.

Another technique of estimating the \Dstst\ \bkgd\ error is to vary the
branching fractions of the P-wave $D$ meson and non-resonant mode components
while keeping the overall peaking \bkgd\ fixed.  The results of using this
technique come out to be of a similar scale to using the above method,
which relies on fits to data.  Using both techniques would
double-count the error, so we rely on the data-fit technique alone to
estimate the error from this source.

\sssec{ \Comb\ Background Shape Uncertainty  }

A procedure similar to that employed for the peaking \bkgd\ shape is
used to estimate the impact of the uncertainty on the shape of the
\combb, except in this case we work with the $\Delta m$ fits rather
than the $\cosby$ fits.  As before, we do this fit in five
equal-sized bins for each \kv\ over the full kinematically allowed
range of the variable, in both data and Monte Carlo.  We then take
data-to-Monte Carlo ratios of the fit results per bin, and find the
results shown in Fig.(\ref{fig:comborat}), which we then can
approximate with linear fits.  We use these fit results with their
associated {\rncversion one-sigma} slope uncertainties shown in
Table~\ref{table:dmslopes} to weight the \combb\ fraction up and down
and fit to the data minus this reweighted \dist.  The \diff\ from the
original fits gives us an estimate of the error due to the shape of
\combb\ uncertainty for each \kv.  Adding the uncertainties for each
\kv\ in quadrature gives us the total error from this source.

\sssec{Background Check with Goodness-of-Fit}

The quality of the fit as measured by the binned $\chi^2$ (see
Sec.~\ref{sec:gof}) is sensitive to the background fraction
assumed. Thus, we can use $\chi^2$ to check our estimate of the
backgrounds. Figure \ref{fig:chisqvsf} shows plots of $\chi^2$ versus
the fractions of the combinatorial, $\dstst$, and other peaking
backgrounds. These scans yield estimates of the background fractions {\rncversion(in percent)}
of

\bea
\label{eq:threefs}
\fcomb=5.4\pm 1.3, \nonumber 
\\
f_\dstst=6.1\pm 1.2,
\\
\fother=6.4\pm 0.9, \nonumber
\eea
which are in good agreement with the values of $5.33\%$, $4.85\%$ and $7.03\%$ obtained from
the $\Delta m$ and \cosby\ fits in section~\ref{sec:back}.

The good agreement indicates not only that the sizes of 
these backgrounds are well-estimated, but
that the shape in the {\rncversion four-dimensional kinematic-variable phase-space} 
given by the Monte Carlo agrees well with their shape in the data.

 \begin{figure}[ht]
 \bcenter    
 {\parbox{10cm}
 {\resizebox{!}{ 15cm}{\includegraphics{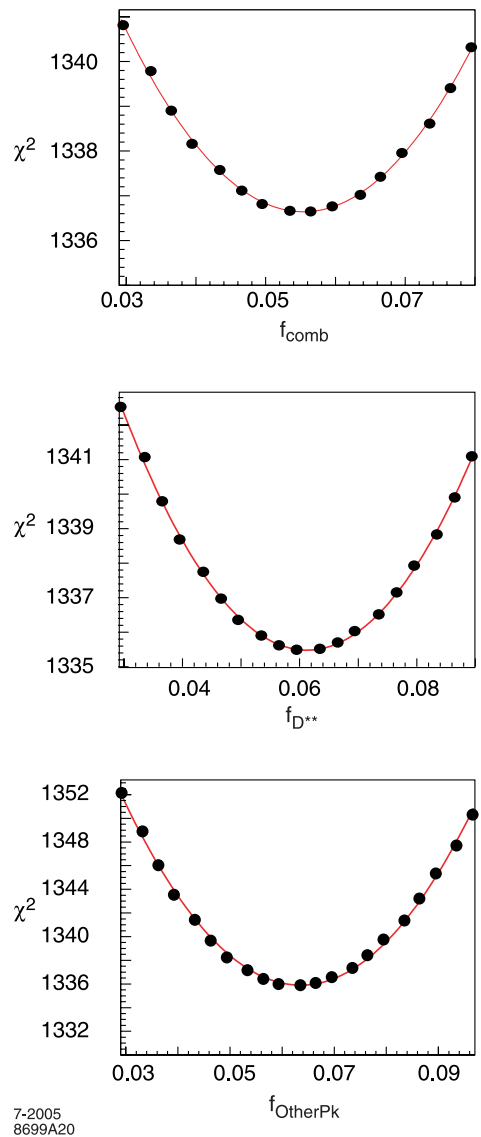}} }
} 
 \caption{ \label{fig:chisqvsf}
$\chi^2$ of fit vs. fractions of combinatorial ($f_{\rm comb}$), 
\dstst ($f_\dstst$) and other
peaking backgrounds ($\fother$).  From this we see that our method of
ascertaining the goodness-of-fit is independently indicating that the
values we obtained for these fractions from the \cosby\ and $\Delta m$
fits were correct.  } 
\ecenter
\end{figure}

\sssec{Total Peaking and Combinatorial Background Fractions}

Another source of error is the uncertainty in the \normzn\ of the
peaking and \comb\ background fractions.  To estimate this error we
vary the fractions $\fcomb$, $f_\dstst$ and $\fother$ by 
their
uncertainties and assess the impact on the fit parameters. The
uncertainties for the combinatorial and \dstst\ fractions are taken
from the $\Delta m$ and \cosby\ fits. However, since $\fother$ is
input to these fits by scaling the Monte Carlo \dist, we take its
uncertainty instead from the goodness-of-fit scan of the previous
section. In varying $\fother$ we keep the total peaking $f_{\rm 
peaking}=f_\dstst+\fother$ fixed, since this behavior is observed
when $\fother$ is varied in the \cosby\ fit.

\sssec{MC/Data Sideband Comparison}

 The distributions of the kinematic variables for Monte Carlo and
data agree well
 in the $\Delta m$ sideband region used to estimate
the combinatorial background. But since the \comb\ \bkgd\ comprises
about a third of the
 total \bkgd\ under the peak, the small
differences in the shapes of
 the distributions could introduce an
error in the background subtraction process. To estimate the impact of
this effect, we first take the data-to-Monte Carlo ratios in the
\sbr\ and then fit these ratios with polynomials.

We next use these functions one at a time to multiply the
combinatorial background from Monte Carlo before it is subtracted from
the data to prepare the sample for fitting.  We then carry through
the fits and take the differences in the \FF s we obtain from these
with the \FF s we obtained from the fits with the unaltered background.
This procedure yields the results shown in Table~\ref{rwtcombbkgd}.
 
The differences are small. The largest is from using the
function for \recow, from which we find $\Delta \rho^2$ $\sim
0.006$. Since in the end we take the $w$-dependence of the \comb\ background
from the data and the deviations due to weighting the angular
distributions are small, we add nothing to the systematic
uncertainty from this check.

\vsp
\begin{table}[ht]
\caption{\label{rwtcombbkgd}  Changes
in the fitted parameters for reweighting of the Monte Carlo combinatorial
background distributions in the four kinematic variables.}
\bcenter
\begin{tabular}{lccc}
\hline \hline
Reweighted distributions & $R_1$ & $R_2$ & $\rho^2$ \\
\hline
$w$ distribution     & $-0.002$ & $\hphantom{+}0.0\hphantom{00}$ & $\hphantom{+}0.006$ \\
\ctl\ distribution & $\hphantom{+}0.001$ & $-0.002$ & $-0.001$ \\
\ctv\ distribution & $\hphantom{+}0.002$ & $-0.003$ & $\hphantom{+}0.001$ \\
\angchi\ distribution & $\hphantom{+}0.004$ & $-0.002$ & $\hphantom{+}0.001$ \\
\hline\hline
\end{tabular}
\ecenter
\end{table}

\subsection{ Summary of Systematic Errors  }

The systematic errors are summarized in Table
\ref{table:CLNsysterrs}.  The largest contributions to the systematic errors of \rone\ and \rtwo\ arise from the uncertainties in \bkgd\ composition, and in the \normzn\ of non-\dstst\ peaking \bkgd s.  On the other hand, the  largest contributions to the systematic error of \rhosq\ arise from the uncertainties in the \kv-dependence of the \bkgd s, in the slow pion tracking \eff\ and also in the \normzn\ of non-\dstst\ peaking \bkgd s.

While the total error remains statistics-dominated, this result is
only about a factor of two above the systematic limit. An improved
understanding of the systematics would be needed if we were to extend
this analysis to other $D$ decay modes and a larger data set.

\begin{table}[ht!]
\caption{\label{table:CLNsysterrs}  Summary of the estimated \syst\ errors for the baseline fit.   }
\begin{center}
\begin{tabular}{ lllll }
\hline\hline
\em{Error source}        &  $\sigma_{\rone}$ & $\sigma_{\rtwo}$ &  $\sigma_{\rhosq}$  \\
\hline
Charged particle track efficiency  & 0.005 & 0.004 & 0.003   \\  
Slow pion track efficiency      & 0.003 & 0.000 & 0.018  \\
PID misID (electron, kaon)        & 0.006 & 0.004  & 0.003  \\
\dstst\ background normalization        & 0.003 & 0.002  & 0.005  \\
Other peaking backgrounds normalization & 0.021   & 0.006    &  0.015   \\
Combinatorial background normalization  & 0.002 & 0.001 & 0.003  \\ 
Bkgd composition (branching fractions)     & 0.016 & 0.008 & 0.005  \\
 \KV-dependence of bkgd    & 0.001 & 0.001  & 0.028  \\  
Final state radiation           & 0.005 & 0.002 & 0.002 \\
\hline 
Total  \Syst\                   &  0.027 & 0.013 &  0.035   \\
\hline\hline
\end{tabular}
\end{center}
\end{table}

\section{Summary and Discussion}
\label{sec:Summary}

Based on a sample of 16,386 \Btodstarenu\ events, we have measured the
form factors $A_1$, $V$ and $A_2$ in terms of the HQET-inspired
parameters $R_1$, $R_2$ and $\rho^2$.  The baseline result, using the CLN
\paramzn\ of \haonew\ (see Sec.~\ref{sec:hqsrships}) is obtained
neglecting any possible $w$-dependence of $R_1(w)$ and $R_2(w)$, and
including all errors is

\bea
\label{eq:results}
R_1=1.396\pm 0.060\pm 0.035\pm 0.027, \nonumber \\
R_2=0.885\pm 0.040\pm 0.022\pm 0.013,  \\
\rho^2=1.145\pm 0.059\pm 0.030\pm 0.035, \nonumber \\
\nonumber
\eea
where the first error is \statl, the second Monte Carlo \statl, and
the third \syst. 

Note that the CLEO result discussed in Sec.~\ref{sec:Introduction} was
obtained using the expansion in $(w-1)$ (see Eq.~(\ref{eq:ffexpd}))
to linear order.  When analyzed in this manner, the current
dataset yields $\rone=1.40\pm 0.06$, $\rtwo=0.87\pm 0.04$, and
$\rhosq=0.79\pm 0.06$. CLEO also obtained $\rho^2$ using the form
(see Eq.(\ref{eq:oliver})) 
suggested by reference \cite{oliver} that is very close numerically to 
the CLN parameterization we use. 
In this case they obtained $\rho^2=1.42\pm 0.32$ which is consistent with
our result.

The values of \rhosq\ obtained are very different using the linear 
parameterization,
as mentioned in Sec.~\ref{sec:Introduction}.  This is
because the linear fit significantly underestimates the slope as it
attempts to accommodate the higher-order terms ({\it i.e.}, at $w =
1$, the absolute value of the slope is larger assuming a curve for the
fit vs. a line).

The $\chi^2$ of this linear fit is $1337.98$ which is $1.32$
units of $\chi^2$ worse than the baseline fit using the expansion of
Eq.~(\ref{eq:clnff}) (see Eq.(~\ref{eq:basechisq})). Thus while the baseline 
fit is slightly better than the linear, both are acceptable.

The sensitivity needed to independently establish the $w$-dependence of
$R_1(w)$ and $R_2(w)$ is not yet available, but theoretical
predictions may be used to make comparisons.  If we compare the
predictions of CLN~\cite{CLN} for $R_1(w)$ and $R_2(w)$ (see
Table~\ref{table:theorydep}) to the baseline result, the shifts are
found to be $\Delta R_1=1.42-1.27=0.15$ ($\approx 2.0\sigma$ apart)
and $\Delta R_2=0.87-0.80=0.07$ ($\approx 1.5\sigma$ apart). For
Ligeti and Grinstein~\cite{LigetiGrinstein} the numbers are
$\Delta\rone=1.42-1.25=0.17$ ($2.3\sigma$) and
$\Delta\rtwo=0.87-0.81=0.06$ ($\approx 1.3\sigma$). The older
prediction of Neubert~\cite{NeubertPhysReport} is closer with $\Delta
R_1 $ giving $\approx 0.6\sigma$.  If the theoretical error on \rone\ is $\sim
0.03$ as estimated by Ligeti~\cite{ligetipc} then there is a
mild indication of disagreement. Higher statistics will be needed to
resolve or confirm the possible discrepancy.

These results allow a five-fold reduction in the largest source of
systematic error in $V_{\rm cb}$ measurements based on $B\to
\Dst\ell\nu$ decays to be made.  Using the measurements of \rone\ and
\rtwo\ presented in this paper, we update the result from
 the 2004 \babar\ paper\cite{babarVcb} to find
\be
|V_{\rm cb}|=37.6\pm 0.3(stat)\pm
1.3(syst)\hphantom{x}^{+1.5}_{-1.3}(theory)
\\
\times 10^{-3}\nonumber. 
\ee
where small correlations between the present analysis and that of
Ref.~\cite{babarVcb} have been ignored.  The error due to the uncertainties in $R_1$ and
$R_2$ is reduced from $\hphantom{x}^{+2.9\%}_{-2.6\%}$ to $\pm 0.5\%$.
The overall systematic error drops from $\pm 1.7 $ to $\pm 1.3 $.  The
other systematic errors remain the same as published in
Ref.~\cite{babarVcb}.

A considerable improvement was also obtained in measurements of the
lepton endpoint spectrum in $b \rar u \ell \nu$ decays. Compared to
usage of the old CLEO $B \rar \Dst$ form-factor measurements, usage of these newer form factors allowed the
systematic error on  the inclusive $B\rar X_u \ell \nu$ branching fraction for decays with a lepton in
the momentum range $2.0-2.6
\gevc$ to be reduced from $6.7\%$ to $2.4\%$. In the higher momentum range 
$2.3-2.6\gevc$ the improvement was from $2.8\% $ to $1.3\% $. This
enabled a significant improvement in the measurement of $|V_{\rm ub}|$
from the endpoint spectrum\cite{ref:endpt}.

In addition we have demonstrated useful approximations to the maximum
likelihood method that allow us to cope with the limited size of the
Monte Carlo samples available to modern high-luminosity
experiments. We have also developed the procedures needed to evaluate
the corrections and additional uncertainties due to these
approximations. These methods are not unique to $B \rar D^*\ell\nu$ or
$B \rar J/\psi K^*$ decays, but could be applied to any analysis
with a complex multi-dimensional acceptance and resolution functions.

An approach to multi-dimensional goodness-of-fit criteria that allows
assessment of the quality of a likelihood fit has also been
developed. A binned $\chi^2$ measure, with the method developed to
estimate the errors, yields a measure of goodness-of-fit that has a
straightforward statistical interpretation which is easily and
intuitively understandable.

\section{Acknowledgments}
\label{sec:Acknowledgments}

The authors wish to thank B.~Grinstein, Z.~Ligeti, M.~Neubert,
L.~Lellouch and L.~Oliver for very useful discussions of the 
theoretical issues 
involved in this work.


We are grateful for the 
extraordinary contributions of our \pep2\ colleagues in
achieving the excellent luminosity and machine conditions
that have made this work possible.
The success of this project also relies critically on the 
expertise and dedication of the computing organizations that 
support \babar.
The collaborating institutions wish to thank 
SLAC for its support and the kind hospitality extended to them. 
This work is supported by the
US Department of Energy
and National Science Foundation, the
Natural Sciences and Engineering Research Council (Canada),
Institute of High Energy Physics (China), the
Commissariat \`a l'Energie Atomique and
Institut National de Physique Nucl\'eaire et de Physique des Particules
(France), the
Bundesministerium f\"ur Bildung und Forschung and
Deutsche Forschungsgemeinschaft
(Germany), the
Istituto Nazionale di Fisica Nucleare (Italy),
the Foundation for Fundamental Research on Matter (The Netherlands),
the Research Council of Norway, the
Ministry of Science and Technology of the Russian Federation, and the
Particle Physics and Astronomy Research Council (United Kingdom). 
Individuals have received support from 
CONACyT (Mexico),
the A. P. Sloan Foundation, 
the Research Corporation,
and the Alexander von Humboldt Foundation.

\clearpage

\clearpage

\appendix{\bf  Appendix A: Evaluation of additional errors}
\label{sec:appxsysterrs}
\def\parta{\frac\partial{\partial\mu_\alpha}}
\def\partb{\frac\partial{\partial\mu_\beta}}
\def\partmu{\frac\partial{\partial\mu}}
\def\dmu {\partial\mu}
\def\intf{\int dx f}
\def\expect#1{\langle #1\rangle}

\vsps

We supply here some of the mathematical details associated with biases
and uncertainties introduced in our application of the maximum
likelihood method.  See Ref.~\cite{Gill} for further details.

Except as noted, in the below, we use the notation of Sec.~\ref{sec:Analysis}.


\subsec{Bias From Approximate PDF}

Suppose events are distributed as 
\be
\frac{{\cal F}(x;\mu_{\rm t})}{\int dx {\cal F}(x;\mu_{\rm_t})}, 
\ee
but the 
unnormalized PDF used in the maximum likelihood fit is $f(x;\mu)$ 
(defined by Eq.(\ref{eq:approx}) and used as an approximation for the unknown
${\cal F}(\tilde x;\mu)$ in Eq.~(\ref{eq:llapprox})).  
As
the number of events $N$ tends to infinity, the sum defining 
the likelihood can be replaced by the integral
\be
\ln L = N\int dx {\cal F}(x;\mu_{\rm t})\ln f(x;\mu) - N\ln \int dx f(x;\mu).
\ee
The second term comes from normalizing the PDF.  If we expand about
$\mu_{\rm t}$ and keep only the leading terms, treating

\be
\frac{{\cal F}(x;\mu_{\rm t})}{\int dx {\cal F}(x;\mu_{\rm t})} -
\frac{f(x;\mu_{\rm t})}{\int dx f(x;\mu_{\rm t})}
\ee
as being itself of order $\mu - \mu_{\rm t}$,
we find that the maximum occurs at the point where the derivatives given by
\begin{widetext}
\be
\footnotesize
\frac\partial{\partial \mu}\ln L =
\int dx {\cal F}(x;\mu_t)  \frac\partial{\partial \mu}\ln f\left(
\frac{{\cal F}(x;\mu_{\rm t})}{\int dx{\cal F}(x;\mu_{\rm t})} -
\frac{f(x;\mu_{\rm t})}{\int dx f(x;\mu_{\rm t})}
\right) 
+\left(\left<{\frac{\partial\ln f}{\partial \mu}}\right>
\left<{\frac{\partial\ln f}{\partial \mu}}\right>^T-
\left<{\left<\frac{\partial\ln f}{\partial \mu}\right>
\left(\frac{\partial\ln f}{\partial \mu}\right)^T}\right>\right)(\mu-\mu_{\rm t})
\ee
are zero. All derivatives are evaluated at $\mu=\mu_{\rm t}$.

When we take the approximation
\be
f(x;\mu)={\cal F}(x;\mu_{\rm mc})\frac{F(x;\mu)}{F(x;\mu_{\rm mc})}
\ee
and expand about $\mu=\mu_{\rm t}$, we find to leading order

\be
\frac{{\cal F}(x;\mu_{\rm t})}{\int dx {\cal F}(x;\mu_{\rm t})} -
\frac{f(x;\mu_{\rm t})}{\int dx f(x;\mu_{\rm t})}
=\frac{{\cal F}(x;\mu_{\rm t})}{\int dx {\cal F}(x;\mu_{\rm t})}
\left[
\expect{\frac{\partial\ln\frac{\cal F}{F}}{\partial\mu}}
-{\frac{\partial\ln\frac{\cal F}{F}}{\partial\mu}}\right]^T(\mu_{\rm mc}-\mu_{\rm t}).
\ee

Setting the derivative of the log-likelihood to zero, we have
\bea
(\mu_{\rm mc}-\mu_{\rm t})\left[
\left<{\frac{\partial\ln\frac{{\cal F}}{F}}{\partial\mu}}\right>
\left<{\frac{\partial\ln F}{\partial\mu}}\right>^T
-\left<{\left(\frac{\partial\ln\frac{{\cal F}}{F}}{\partial\mu}\right)
\left(\frac{\partial\ln F}{\partial\mu}\right)^T
}\right>\right]^T= \nonumber\\
-(\mu -\mu_{\rm t})
\left[
\left<{\frac{\partial\ln F}{\partial\mu}}\right>
\left<{\frac{\partial\ln F}{\partial\mu}}\right>^T
-\left<{\left(\frac{\partial\ln F}{\partial\mu}\right)
\left(\frac{\partial\ln F}{\partial\mu}\right)^T
}\right>\right]^T.
\eea
\end{widetext}
Solving for the bias $\mu-\mu_{\rm t}$ yields the result Eq.(\ref{eq:bias}).

\subsec{Error from Monte Carlo evaluation of normalization}

The error induced by the use of Monte Carlo integration to normalize a PDF
can be calculated considering the asymptotic form
\be
\ln L = N\int dx f(x;\mu_{\rm t})\ln f(x;\mu) - N\ln [\int f(x;\mu) dx +\epsilon(\mu)]
\ee
where now we are no longer concerned with the distinction between ${\cal F}$ and $f$.  The function $\epsilon(\mu)$ is the deviation of the Monte Carlo 
integral from the true integral.  The maximization equation now reads
\begin{widetext}
\be
\label{eq:biasfind}
\frac\partial{\partial \mu}\ln L =
\left(\left<{\frac{\partial\ln f}{\partial \mu}}\right>
\left<{\frac{\partial\ln f}{\partial \mu}}\right>^T
\left<{\left(\frac{\partial\ln f}{\partial \mu}\right)
\left(\frac{\partial\ln f}{\partial \mu}\right)}\right>^T\right)(\mu-\mu_{\rm t})
-\frac\partial{\partial\mu}\frac{\epsilon(\mu)}{\int dx f(x;\mu)},
\ee
so, again setting the log of the likelihood to zero, the bias is
\be
(\mu-\mu_{\rm t})=-\frac\partial{\partial\mu}\frac{\epsilon(\mu)}{\int dx f(x;\mu)} E,
\ee
where we recognize that the term multiplying $(\mu-\mu_t$) in 
Eq.(\ref{eq:biasfind}) is the
inverse of the error matrix $E$.

Now we suppose the Monte Carlo simulation to be accurate so there is no
bias, but an uncertainty is introduced by the fluctuations in $\epsilon$.  We therefore need an estimate of the matrix
\be
\left<{\left(\frac\partial{\partial\mu_\alpha}\frac{\epsilon(\mu)}{\int dy f(y;\mu)}\right)
\left(\frac\partial{\partial\mu}\frac{\epsilon(\mu)}{\int dy' f(y';\mu)}\right)^T}\right>.
\ee

A straightforward calculation gives

\bea
&&\frac 1{(\intf)^2}\int dx \left(\frac{\partial f}{\partial\mu}\right) \left(\frac{\partial f}{\partial\mu}\right)^T
\hfill\nonumber\\&&
-\frac 1{(\int dx f)^3}\left[
\left(\int dx\frac{\partial f}{\partial\mu}\right)
\left(\int dx f\frac{\partial f}{\partial\mu} \right)^T  +
\left(\int dx f\frac{\partial f}{\partial\mu}\right)
\left(\int dx \frac{\partial f}{\partial\mu} \right)^T\right]
\nonumber\\&&\qquad  \qquad \qquad  \qquad +
\frac 1{(\int dx f)^4} \left(\int dx \frac{\partial f}{\partial\mu}\right)
\left(\int dx \frac{\partial f}{\partial\mu}\right)^T\int dx f^2.
\eea

To apply this to our circumstance where we perform a Monte Carlo integration
with points having the density ${\cal F}(x;\mu_{\rm mc})$, we need to make the
identifications
\be
\frac 1{N_{MC}}\sum_i\ldots \to \frac 1{\int dx {\cal F}}\int dx {\cal F}(x;\mu_{\rm mc})\ldots;\quad f\to \frac{F(x;\mu)}{F(x;\mu_{\rm mc})}.
\ee
In the final expressions, we do not need to make the distinctions between
$\mu$, $\mu_{\rm mc}$, and $\mu_{\rm t}$ which is only necessary when taking the derivatives
with respect to $\mu$.  These differences are of higher order.  This results in the correspondences
\bea
\int dx f &\to& \int dx {\cal F}(x;\mu_{\rm t})\\
\int dx \frac{\partial f}{\partial\mu} &\to& \int dx  {\cal F}(x;\mu_{\rm t})\frac{\partial\ln F(x;\mu)}{\partial\mu}
=\left<{\frac{\partial\ln F}{\partial\mu}}\right>\int dx {\cal F}, {\rm etc.}\\
\eea
where the derivatives are evaluated at $\mu_{\rm t}$

In this way we find 
\bea
\left<{\left(\frac\partial{\partial\mu}\frac{\epsilon(\mu)}{\int dy f(y;\mu)}\right)
\left(\frac\partial{\partial\mu_\beta}\frac{\epsilon(\mu)}{\int dy' f(y';\mu)}\right)^T}\right>
&=& \frac {1}{M_{MC}}
\left<{\left(\frac{\partial\ln F}{\partial\mu}\right)
\left(\frac{\partial\ln F}{\partial\mu}\right)}\right>^T
-\left<{\frac{\partial\ln F}{\partial\mu}}\right>
\left<{\frac{\partial\ln F}{\partial\mu}}\right>^T
\nonumber\\
&=&  \frac 1{M_{MC}}E^{-1}
\eea
and
\be
\left<{\left(\mu-\mu_{\rm t}\right)
\left(\mu-\mu_{\rm t}\right)^T}\right>
=E\frac 1{M_{MC}}E^{-1}E=
\frac 1{M_{MC}}E,
\ee
\end{widetext}
where $M_{MC}=N_{MC}/N_{data}$ is the ratio between the number of MC events 
and the number of signal events in the data (the inverse of the $\rmc$  used in Sec.~\ref{sec:gof}).
That is to say, the error matrix due to Monte Carlo integration has the 
same form as that for the statistical error from data, 
except that it is $1/N_{MC}$
instead of $1/N_{\rm data}$ that enters.

\subsec{ Pseudo-log-likelihood Error}

The pseudo-log-likelihood error can be computed from sums over the Monte Carlo sample
used in the background subtraction.  In this case the weights ($w_i$)
are those used to weight each type of background to obtain the correct
normalization as described in Sec. \ref{sec:back} (see Eq.~(\ref{eq:bkwt})).

We consider a pseudo-log-likelihood
\bea
\ln\Lambda &=&\sum_{i(S)} \ln F_S(x_i;\mu) 
+ \sum_{j(B)} \ln F_S(x_j;\mu)\nonumber\\&&-\sum_{j'(B,MC)} \ln F_S(x_{j'};\mu)-N_S\int dx F_S(x;\mu)\nonumber\\
\eea
where the $S$, $B$, and $(B,MC)$ indicate that the sums are over signal events,
background events, and background MC events, respectively.  Of course we cannot know event-by-event which are signal and which are background events, but 
for the sum this doesn't matter.  The signal distribution is indicated by
$F_S(x;\mu)$; the background is $F_B(x)$. We expand about $\mu=\mu_{\rm t}$ and set
$(\partial/\partial\mu_\alpha)\ln\Lambda$ to zero to find 
the background subtraction error $E_S$ is given by
\bea
N_S E_S^{-1}(\mu-\mu_{\rm t})&=&
\sum_{i(S)}\partmu\delta\ln F_S(x_i;\mu)\nonumber\\
&&+\sum_{j(B)}\partmu\delta\ln F_S(x_j;\mu)\nonumber\\
&&\qquad
-\sum_{j'(B,MC)}\frac{N_B}{N_{MC}}\partmu\delta\ln F_S(x_{j'};\mu)\nonumber\\
\eea
where
\be
\delta\ln F_S(x_i;\mu)=\ln F_S(x_i;\mu)
-\frac{\int dx F_S(x;\mu_{\rm t})\ln F_S(x;\mu)}{\int dx F_S(x;\mu_{\rm t})},
\ee
and

\bea
E_S^{-1}=\left<{\left(\frac{\partial\ln F_S}{\partial\mu}\right)
\left(\frac{\partial\ln F_S}{\partial\mu}\right)^T
}\right>_S
\nonumber\\
-\left<{\frac{\partial\ln F_S}{\partial\mu}}\right>_S
\left<{\frac{\partial\ln F_S}{\partial\mu}}\right>_S^T
\eea

where 

\be
\left<{A }\right>_S=\int dx F_S(x;\mu_t)A(x)/\int dx F_S(x;\mu_{\rm t})
\ee

The fluctuations average to zero, but their squares do not.  For example,
\begin{widetext}
\bea
& &\sum_{i(S)}\sum_{i'(S)}\left(\partmu\delta\ln F_S(x_i;\mu_{\rm t})\right)
\left(\partmu\delta\ln F_S(x_{i'};\mu_{\rm t})\right)^T
= \\
& & N_S\left[\left<{\left(\frac{\partial\delta\ln F_S}{\dmu}\right)
\left(\frac{\partial\delta\ln\ F_S}{\dmu}\right)}\right>_S^T 
-\left<{\frac{\partial\delta\ln F_S}{\dmu}}\right>_S
\left<{\frac{\partial\delta\ln F_S}{\dmu}}\right>_S^T\right];
\nonumber\\
& &\sum_{j(B)}\sum_{j'(B)}\left(\partmu\delta\ln F_S(x_j;\mu_{\rm t})\right)
\left(\partmu\delta\ln F_S(x_{j'};\mu_{\rm t})\right)^T\nonumber
= \\
& & N_B\left[\left<{\left(\frac{\partial\delta\ln F_S}{\dmu}\right)
\left(\frac{\partial\delta\ln F_S}{\dmu}\right)}\right>_B^T 
-\left<{\frac{\partial\delta\ln F_S}{\dmu}}\right>_B
\left<{\frac{\partial\delta\ln F_S}{\dmu}}\right>_B^T\right]
\eea
\end{widetext}
where now
\be
\expect{A }_B=\int dx F_B(x;\mu_t)A(x)/\int dx F_B(x;\mu_{\rm t}).
\ee
Altogether we find
\bea
&&\left<{\left(\mu-\mu_{\rm t}\right) \left(\mu-\mu_{\rm t}\right)^T}\right>
=\frac 1{N_S}{E_S}\nonumber\\
&&\qquad+\frac1{N_S}\frac{N_B}{N_S}\left(1+\frac{N_B}{N_{MC}}\right){E_S} {E_B}^{-1}{E_S}.\nonumber\\
\eea

\appendix{ \bf Appendix B: Background estimation}
\label{sec:appxbkgd}

\vsps

This appendix gives more detail on the background estimation
procedures.

\subsec{ Combinatorial background}

Figure~\ref{fig:dmall1} shows the distribution of \deltam\ for the
data sample on a linear scale.
The linear scale is used so the level of \comb\ \bkgd\ can be judged by
eye from the tail extending outwards towards the right (the peaking \bkgd s are
underneath the signal peak here and their levels cannot be
distinguished from the data). The combinatorial background appears to be of order
$5\%$, but it is hidden under a signal that has substantial tails. It
is difficult to distinguish tail from background.  To extract the
background fraction and estimate the signal  this
distribution is fit to a set of Gaussians representing the signal and to a
threshold function to model the background.  To investigate the
background shape and the tails of the signal \MC\ simulated events are used.

\begin{figure}[htp]
\begin{center}
{\parbox{8cm}
  {\resizebox{8cm}{!}{\includegraphics{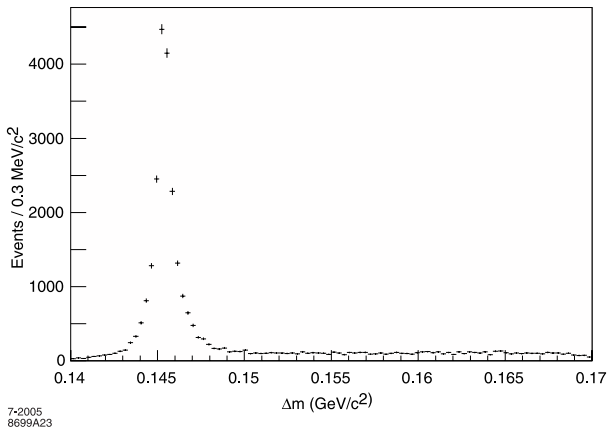}}
}}
\end{center}
\caption{
The \deltam\ data distribution for $\dst e\nu$ candidates on a linear scale.
}
\label{fig:dmall1}
\end{figure}

First, to investigate the shape needed to describe the background, a pure combinatorial background sample selected from the \MCs\ is used.

Figure~\ref{fig:dmcombo} shows the distribution of \MC\ combinatorial
background events fit to the (a) unextended (b) and extended threshold
function, as discussed in Sec.~\ref{sec:combobkgd}. Though not
immediately obvious by eye, the unextended function systematically
underestimates the data in the signal region near $\deltam=0.1454
\gevcc$. The true number in the signal box ($0.143\leq\deltam\leq
0.148 \gevcc$) is 3540. The unextended fit yields $3400\pm 39$ whereas the
extended fit yields the much closer result $3528\pm 52$. Fits to
subsets of the Monte Carlo data ({\it e.g.}, those binned in the kinematic
variables) show the same pattern with the fit to the extended function
having
much better success at estimating the number of background events than the
unextended.

Next, Figure~\ref{fig:dmsignal}a shows the \deltam\ distribution of a \MC\ simulated
pure \dstar\ signal sample. 
There are substantial tails and even the
core is not consistent with a single Gaussian. It takes five Gaussians 
to achieve the good fit shown in Figure~\ref{fig:dmsignal}a.

The fit to the full MC \deltam\ distribution is finally shown in 
Figure~\ref{fig:dmsignal}b. 
In this fit the three core Gaussians of the 
signal are allowed to  float, but the tail is fixed. For the threshold function{\rncversion ,} the threshold 
and the scale factor are fixed while the other two parameters are fitted.
This procedure reproduces the \MC\ input signal and background fractions.


The \deltam\ distribution of the data with the fit as described is
shown in Figure~\ref{fig:deltamdata}. We find $N^{fit}_{cb}=1325\pm 65$
which corresponds to a combinatorial background fraction of $6.4\pm
0.3\%$ before cutting on \cby.

To check the stability of this fit{\rncversion ,} the fitting
conditions are varied. Letting the fourth Gaussian float yields $1374\pm 92$. The
error on the difference is $\pm 65$, so this is consistent with the
fixed fourth Gaussian used for the central value.  Moving the long
tail represented by the fifth Gaussian up and down by $50\%$ results in a
variation of $\approx\pm26$ in the background estimate. This
variation is added in quadrature with the error reported by the fit to yield a
total error of $\pm 70$. The combinatorial background fraction with error 
becomes $6.39\pm0.32\%$.

\begin{figure*}[htp]
\begin{center}
{\parbox{8cm}
  {\resizebox{8cm}{!}{
  \includegraphics{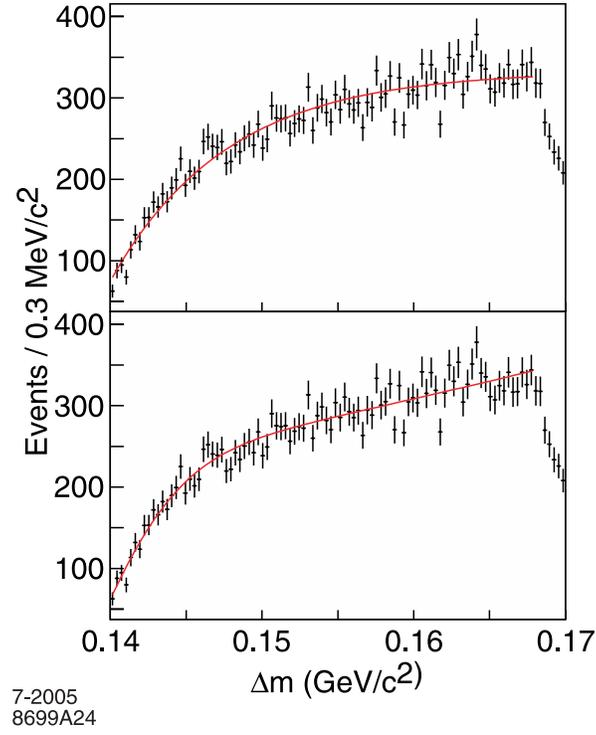}
  }
}}
\end{center}
\caption{\label{fig:dmcombo} The \deltam\ distribution for combinatorial  
background selected from the Monte Carlo with fit to the unextended threshold 
function (a) and to the extended function. 
}
\end{figure*}

\begin{figure*}[htp]
\begin{center}
{\parbox{8cm}
  {\resizebox{8cm}{!}{
  \includegraphics{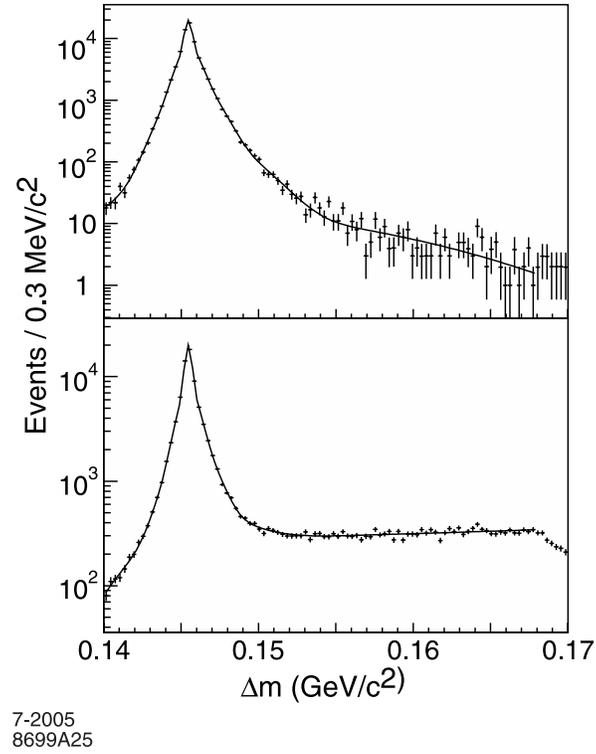}
  }
}}
\end{center}
\caption{\label{fig:dmsignal} The \deltam\ distribution for (a) pure \dstar\ signal selected from the
Monte Carlo with fit to five Gaussians and (b) the \deltam\
distributions on the full \MC\ sample, fit to the five Gaussian signal
shape plus extended threshold function.
}
\end{figure*}

\subsec{ Fits in \kv\ bins}

To estimate the difference in {\rncversion kinematic-variable} dependence between
data and MC, bin-by-bin fits in five bins for each of the four \kv
s are done.  In the five-bin fits for the {\rncversion combinatorial-background kinematic-variable} dependence (as
discussed in Sec.~\ref{sec:combobkgd}) it is found that only four
Gaussians are needed to represent the signal shape and that the $\pm 50\%$
variation of the tail Gaussian produces negligible changes in the
estimated background level. The \deltam\ plots with fits are shown in
Figure~\ref{fig:dmfitbins}. The data-to-\MC\ ratio plots which result
from these fits is shown
{\rncversion Fig.}~\ref{fig:comborat} 
in Section~\ref{sec:Analysis}.

\begin{figure*}[htp]
\begin{center}

{\parbox{16cm}
  {\resizebox{16cm}{!}{
  \includegraphics{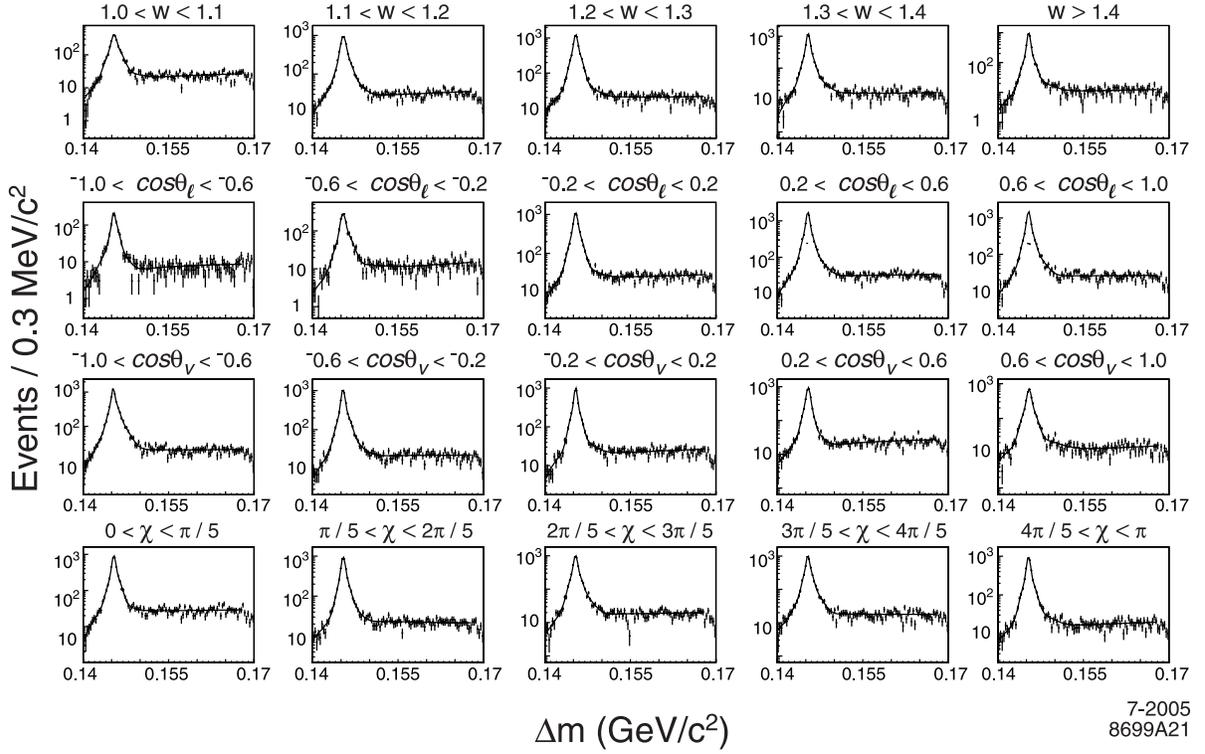}
  }
} 
}
\end{center}
\caption{\label{fig:dmdata} The \deltam\ distribution for data in kinematic variable bins. The rows are fits
for $w$, \ctl, \ctv and $\chi$ in each of the five bins spanning the
full range of the variable.
}
\label{fig:dmfitbins}
\end{figure*}

\subsec{ Peaking background including \dstst\ \bkgd\ fits.}

Similarly,  \cosby\ fits in five bins for  each of the four \kv s are 
done to extract
the shape of the \dstst\ \bkgd.
The results of each fit are shown in Figure~\ref{fig:cbybins}. The data-to-\MC\
ratio used to weight the background obtained from these fits are
shown in Figure~\ref{fig:cbyslopes} in Section~\ref{sec:Analysis}.

\begin{figure*}[htp]
\begin{center}
{\parbox{16cm}
  {\resizebox{16cm}{!}{
  \includegraphics{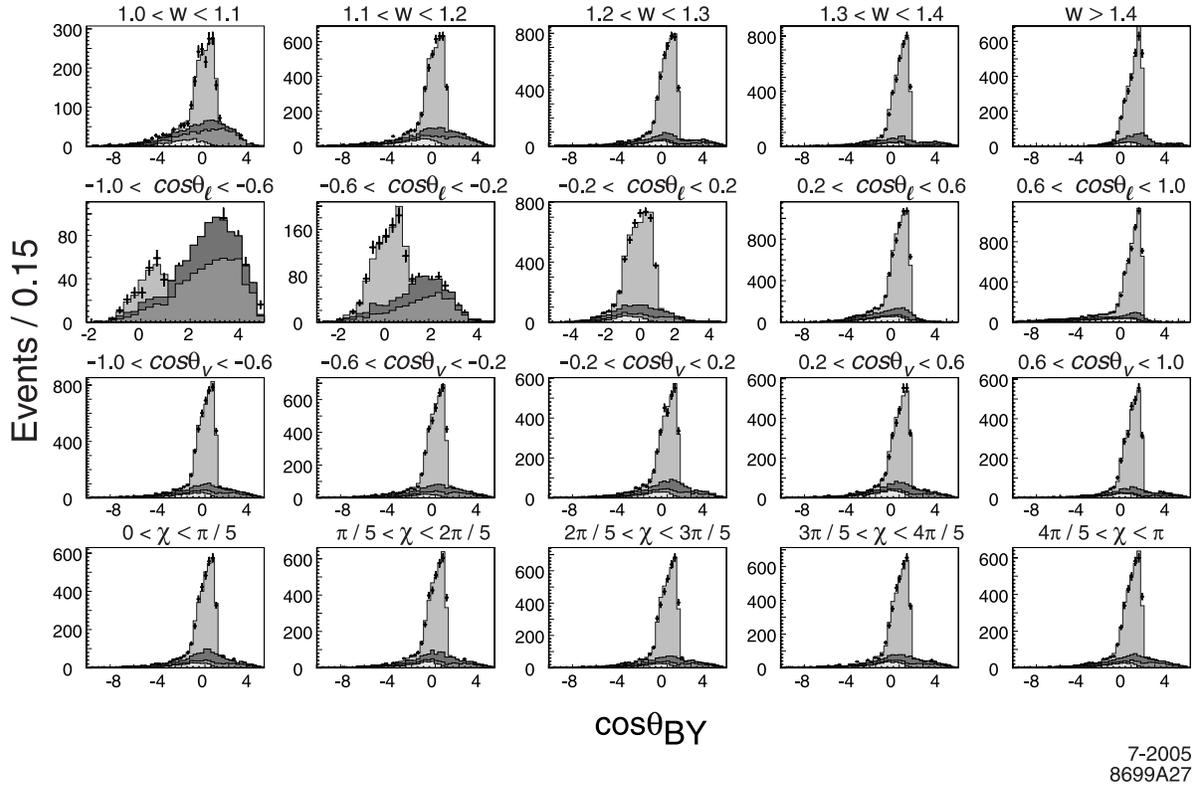}
  }
} 
}
\end{center}
\caption{\label{fig:cbybins} The \cby\ distributions fit to background plus signal in five bins in
each of the kinematic variables. Each row shows the plots for the five
bins corresponding to the variables $w$, \ctl,
\ctv and $\chi$.  The shading is the same as that in Figure~\ref{fig:cbyslopes}.
}
\end{figure*}


\begin{thebibliography}{99}


\bibitem{CLEO} { J.E.~Duboscq {\it et.al.} (CLEO Collaboration), Phys.\
Rev.\ Lett.  {\bf 76},  3898 (1996).}

\bibitem{NeubertPhysReport} For a theoretical review, see, for example, 
{ M.~Neubert,
  ``Heavy quark symmetry,''
  Phys.\ Rep.\  {\bf 245}, 259 (1994)
  [arXiv:hep-ph/9306320].
}

\bibitem{richmanburchat} {For a review of semileptonic $B$ decays, see for
example, J.~D.~Richman and P.~R.~Burchat, ``Leptonic and semileptonic
decays of charm and bottom hadrons,'' Rev.\ Mod.\ Phys.\ {\bf 67}, 893
(1995) [arXiv:hep-ph/9508250]. } 

\bibitem{ARGUS} {H.~Albrecht {\it et. al.} (ARGUS Collaboration),
Z.~Phys. {\bf C 57}, 533 (1993). }



\bibitem{ALEPH1} D.~Buskulic {\it et. al.} (ALEPH { Collaboration}), 
Phys.~Lett. {\bf B 359}, 236 (1995).

\bibitem{ALEPH2} D.~Buskulic {\it et. al.} (ALEPH { Collaboration}), 
Phys.~Lett. {\bf B 395}, 373 (1997).


\bibitem{DELPHI} P.~Abreu {\it et. al.} (DELPHI { Collaboration}), 
Phys.~Lett. {\bf B 510}, 55 (2001).

\bibitem{OPAL} K.~Ackerstaff {\it et. al.} (OPAL { Collaboration}), 
Phys.~Lett. {\bf B 395}, 128 (1997).


\bibitem{babarVcb}
{The \babar\ Collaboration, B.\ Aubert {\em et al.}, 
Phys.\ Rev.\ D-RC {\bf 71}, 051502 (2005) [arXiv: hep-ex/0408027].}

\bibitem{CLN}
{
I.~Caprini, L.~Lellouch and M.~Neubert,
  ``Dispersive bounds on the shape of $\Bbartodslnu$ form
  factors,''
  Nucl.\ Phys.\ B {\bf 530}, 153 (1998)
  [arXiv:hep-ph/9712417].}

\bibitem{GrinsteinLebed}
{
C.~G.~Boyd, B.~Grinstein and R.~F.~Lebed,
  ``Model independent determinations of $\Bbartodslnu$ 
form-factors,''
  Nucl.\ Phys.\ B {\bf 461}, 493 (1996)
  [arXiv:hep-ph/9508211].
}
\bibitem{oliver}
{
A.~Le Yaouanc, L.~Oliver and J.~C.~Raynal, ``Bounds on the derivatives of the Isgur-Wise 
function from sum rules in the heavy quark limit of QCD,''
Phys.\ Lett.\ {\bf B557}, 207 (2003)
  [arXiv:hep-ph/0210231];
A.~Le Yaouanc, L.~Oliver and J.~C.~Raynal,
  ``Sum rules in the heavy quark limit of QCD,''
  Phys.\ Rev.\ D {\bf 67}, 114009 (2003)
  [arXiv:hep-ph/0210233].
}



\bibitem{isgurwise}
{N.~Isgur and M.~B.~Wise,
  ``Weak transition form-factors between heavy mesons,''
  Phys.\ Lett.\ B {\bf 237}, 527 (1990).
}


\bibitem{LigetiGrinstein}
{
B.~Grinstein and Z.~Ligeti,
  ``Heavy quark symmetry in $\Bbartodslnu$ spectra,''
  Phys.\ Lett.\ B {\bf 526}, 345 (2002)
  [Erratum-ibid.\ B {\bf 601}, 236 (2004)]
  [arXiv:hep-ph/0111392].
}

\bibitem{CloseWambach}
{
 F.~E.~Close and A.~Wambach,
  ``Quark model form-factors for heavy quark effective theory,''
  Nucl.\ Phys.\ B {\bf 412}, 169 (1994)
  [arXiv:hep-ph/9307260].
}


\bibitem{ref:babar}
{The \babar\ Collaboration, B.\ Aubert {\em et al.},
Nucl.\ Instrum.\ Methods A {\bf 479}, 1-116 (2002). }


\bibitem{babarsim}
{S.~Agostinelli {\it et. al.}, Geant4 Collaboration, Nucl.\ Instrum.\ 
Methods A {\bf 506}, 250-302 (2003).}

\bibitem{evtgen}
{D.~Lange, Nucl.\ Instrum.\ Methods A {\bf 462}, 152-155 (2001). }



\bibitem{Ryd}
{A.~Ryd, Ph.D. Thesis, University of California, Santa Barbara (UCSB), 
UMI-97-04221-mc (microfiche), (1996).}


\bibitem{Was}                                                                                             
{     S.~Jadach, B.~F.~L.~Ward and Z.~Was,
     ``The precision Monte Carlo event generator KK for two-fermion final  states in e+ e- collisions,''
     Comput.\ Phys.\ Commun.\  {\bf 130}, 260-325 (2000)
   [arXiv:hep-ph/9912214].
}


\bibitem{WasII}
{ E.~Richter-Was, ``QED bremsstrahlung in semileptonic B and leptonic
tau decays,'' Phys.\ Lett.\ B {\bf 303}, 163 (1993).  }


\bibitem{pdg2004}
{Review of Particle Physics,
S.~Eidelman {\it et. al.}, Phys.\ Lett.\ B {\bf 592}, 1 (2004)
[http://pdg.lbl.gov]. }



\bibitem{psikstar}{ ``Measurement of the $B \rar J /\psi K^*(892)$ Decay
Amplitudes,'' by the \babar\ Collaboration, Phys.\ Rev.\ Lett. {\bf 87}, 241801 (2001)
[arXiv:hep-ex/0107049]. } 


\bibitem{Gill} {M.~S.~Gill, Ph.D. Thesis, University of California,
Berkeley (UCB), (2004)  [SLAC Report 794: http://www.slac.stanford.edu/cgi-wrap/pubpage?SLAC-R-794]. }

\bibitem{BbrPseudoL}
{ The \babar\ Collaboration, B.\ Aubert {\em et al.}, 
  ``Ambiguity-free measurement of $\cos(2\beta)$: Time-integrated and
  time-dependent angular analyses of $B $\to$ J/\psi K \pi$,''
  Phys.\ Rev.\ D {\bf 71}, 032005 (2005).
}


\bibitem{bandb}
{
  R.~J.~Barlow and C.~Beeston,
  ``Fitting using finite Monte Carlo samples,''
  Comput.\ Phys.\ Commun.\  {\bf 77}, 219 (1993).
}



\bibitem{oliver2}
{
A.~Le Yaouanc, L.~Oliver and J.~C.~Raynal,
  ``Lower bounds on the curvature of the Isgur-Wise function,'' 
  Phys.\ Rev.\ D {\bf 69}, 094022 (2004)
  [arXiv:hep-ph/0307197].
}


\bibitem{chernoff}
{
H.~Chernoff and E.~L.~Lehman,
Ann.\ Math.\ Stat. {\bf 25}, 579-586 (1954).
}

\bibitem{ligetipc}
{
Z. Ligeti, private communication, May 2005.
}

\bibitem{ref:endpt}
{  The \babar\ Collaboration, B.\ Aubert {\em et al.},  
  ``Measurement of the inclusive electron spectrum in charmless semileptonic B
  decays near the kinematic endpoint and determination of $|$V(ub)$|$,'' to be published in Phys.\ Rev.\ D {\bf 73}, 012066 (2006)
  [arXiv:hep-ex/0509040].
}


\bibitem{ichepLong} { ``Measurement of $B \to D^*$ Form Factors in the Semileptonic Decay $B^{0}\to D^{*-}\ell^+ \nu$'', by the \babar\ Collaboration, submitted
to ICHEP 2004 [arXiv:hep-ex/0409047]. }

\end{thebibliography}
\end{document}